\documentclass[preprints,article,accept,moreauthors,pdftex]{Definitions/mdpi} 

\usepackage{xspace}
\usepackage{bm}
\usepackage{Definitions/aas_macros}
\usepackage{amsmath,amssymb}
\usepackage{upgreek}

\newcommand{\degr}{^\circ}
\newcommand{\F}{\textit{Fermi}\xspace}

\newcommand{\nhd}{$N(\mathrm{H_2})$\xspace}
\newcommand{\wco}{$W_\mathrm{CO}$\xspace}
\newcommand{\hi}{$\mathrm{H\,\scriptstyle{I}}$\xspace}
\newcommand{\hii}{$\mathrm{H\,\scriptstyle{II}}$\xspace}
\newcommand{\hd}{$\mathrm{H}_2$\xspace}
\newcommand{\xco}{$X_\mathrm{CO}$\xspace}
\newcommand{\qhi}{$q_{\mathrm{H\,\scriptstyle{I}}}$\xspace}

\firstpage{1} 
\pubvolume{1}
\issuenum{1}
\articlenumber{0}
\pubyear{2021}
\copyrightyear{2020}
\datereceived{} 
\dateaccepted{} 
\datepublished{} 
\hreflink{https://doi.org/} 

\Title{Gamma rays as probes of cosmic-ray propagation and interactions in galaxies}

\TitleCitation{Gamma rays as probes of cosmic-ray propagation and interactions in galaxies}

\Author{Luigi Tibaldo $^{1,}$*\orcidA{}, Daniele Gaggero $^{2}$, and Pierrick Martin $^{1}$}

\AuthorNames{Luigi Tibaldo, Daniele Gaggero, and Pierrick Martin}

\AuthorCitation{Tibaldo, L.; Gaggero, D.; Martin, P.}

\address{%
$^{1}$ \quad IRAP, Universit\'e de Toulouse, CNRS, UPS, CNES, F-31028 Toulouse, France\\
$^{2}$ \quad Instituto de F\'isica T\'eorica UAM-CSIC, E-28049 Madrid, Spain}

\corres{Correspondence: luigi.tibaldo@irap.omp.eu (L.T.)}

\abstract{Continuum gamma-ray emission produced by interactions of cosmic rays with interstellar matter and radiation fields is a probe of non-thermal particle populations in galaxies. After decades of continuous improvements in experimental techniques and an ever-increasing sky and energy coverage, gamma-ray observations reveal in unprecedented detail the properties of galactic cosmic rays. A variety of scales and environments are now accessible to us, from the local interstellar medium near the Sun and the vicinity of cosmic-ray accelerators, out to the Milky Way at large and beyond, with a growing number of gamma-ray emitting star-forming galaxies. Gamma-ray observations have been pushing forward our understanding of the life cycle of cosmic rays in galaxies and, combined with advances in related domains, they have been challenging standard assumptions in the field and have spurred new developments in modelling approaches and data analysis methods. We provide a review of the status of the subject and discuss perspectives on future progress.}


\begin{document}
\tableofcontents

\section{Context and scope of the review}\label{sec:intro}
Cosmic rays (CRs) are energetic particles first observed around the Earth with energies ranging from MeV to above $10^{20}$~eV and with approximately isotropic arrival directions. They are composed mainly of completely ionised nuclei, with protons accounting for a total fraction $> 90\%$ at GeV energies. They also include electrons, positrons, and antiprotons. The overall CR spectrum follows an approximate power-law distribution, which attests to the non-thermal origin of the particles. A most remarkable change of the power-law spectral slope occurs around $10^{15}$~eV, the so-called knee of the CR spectrum. Below the knee, the standard paradigm holding since the sixties \cite{1964ocr..book.....G} states that the particles originate in the Milky Way, very likely from shock acceleration in supernova remnants (SNR), and diffuse on turbulent magnetic fields in a kpc-sized halo encompassing the disk of the Galaxy for durations exceeding one Myr (see \citet{Gabici:2019jvz} for a recent critical review).

CRs interact with interstellar matter and fields, producing secondary particles and radiation that are indirect means to study CRs in distant locations of the Milky Way, as well as in other galaxies. These observables usefully complement direct measurements of CRs in the heliosphere and allow us to develop our understanding of CR propagation and interactions. Among these indirect probes we find continuum gamma-ray emission produced by inelastic nucleon-nucleon collisions, Bremsstrahlung of CR electrons and positrons interacting with matter, and inverse-Compton (IC) radiation by CR electrons and positrons scattering off low-energy photons. 

Overall, such probes show a fairly good agreement with the standard CR paradigm. However, many aspects are still debated or even largely uncertain, including the range of relevant transport and interaction mechanisms, their uniformity within galaxies, if and how they change based on galactic conditions, the microphysics foundation of all of these aspects, and the role played by CRs in galactic ecosystems. For the latter aspect, let us just briefly mention that CR astrophysics is getting increasing attention for its importance to other astronomical disciplines: astrochemistry and star formation \cite[for a review see][]{Padovani:2020}, galaxy formation and evolution \citep[e.g.,][]{Chan:2019,buck2020,hopkins2021}, and astrobiology \cite[e.g.,][]{globus2020}. 

As early as in 1952 \citet{hayakawa1952} had predicted that the decay of $\pi^0$ produced in inelastic nucleon-nucleon collisions in the Galactic disk would produce a measurable gamma-ray flux. This was confirmed in the '60s and the '70s thanks to the \textit{OSO-3} \cite{clark1968} and \textit{SAS-II} \cite{fichtel1978} satellites, which detected a gamma-ray signal associated with Galactic interstellar matter. The breakthrough in the field came thanks to the \textit{COS-B} satellite (1975-1982), whose observations in the 50 MeV-5 GeV energy range enabled a detailed study of the correlation of gamma-ray emission with interstellar medium (ISM) tracers and provided the first measurements of the large-scale distribution of CRs in the Milky Way \cite{bloemen1989}.

The \textit{Compton Gamma-Ray Observatory} (\textit{CGRO}, 1992-1999) fully covered the energy range 1 MeV-30 GeV thanks to its two instruments COMPTEL and EGRET. \textit{CGRO} data led to many in-depth studies of CRs in the Milky Way. EGRET also first probed CRs in external galaxies in gamma rays by detecting emission from the Large Magellanic Cloud (LMC) \cite{sreekumar1992} and by setting an upper limit on emission from the Small Magellanic Cloud (SMC) \cite{sreekumar1993}. The latter provided an upper limit on GeV CR densities at one third of the value observed locally in the Milky Way, and therefore established observationally the galactic origin of the particles as suggested twenty years earlier by \cite{ginzburg1973}.

The twenty-first century brought numerous and rapid advances in the domain of continuum interstellar gamma-ray emission studies:
\begin{itemize}
\item a dramatic extension of energy interval of the observations, spanning from sub-MeV to sub-PeV energies, with coverage of large portions of the Galactic plane and of the sky;
\item full-sky observations of highest quality in the sub-GeV to sub-TeV domain, most notably thanks to the Large Area Telescope (LAT) onboard the \textit{Fermi Gamma-ray Space Telescope};
\item impressive developments in theoretical and numerical tools for calculating interstellar gamma-ray emission properties, informed by spectacular improvements in the accuracy of direct CR measurements.
\end{itemize}
The next decade holds the promise for further observational advances thanks to new facilities already in the making or still in the planning/proposal phase, with guaranteed steps forward to be made in the TeV domain.

The rapid advances of the past few years and upcoming facilities make it timely to have a new review focussing on gamma rays as probes of CR propagation and interactions in galaxies, as an update to the previous ones touching these subjects \cite{bloemen1989,Strong:2007nh,grenier2015,Gabici:2019jvz}. In Section~\ref{sec:obsmod} we will briefly summarize the status of observational techniques, CR transport theory, modelling and data analysis tools, as well as complementary multiwavelength and multimessenger observations necessary to interpret the gamma-ray data. The following sections will review recent observations, their implications for CR physics, and future perspectives. They are organized around four broad questions.
\begin{description}
\item[Section~\ref{sec:locemiss}] What do we learn from observations of the local interstellar medium near the Sun and how can we use them to connect direct and indirect CR measurements?
\item[Section~\ref{sec:largescale}] What does interstellar emission tell us about the large-scale distributions of CRs in galaxies and what  does it teach us about CR transport?
\item[Section~\ref{sec:youngcr}] What do we know about particle propagation and interactions in the vicinities of sources and what role does this phase play in the CR life cycle?
\item[Section~\ref{sec:galaxies}] What are the properties of gamma-ray emitting galaxies as a population and what do we learn about the variety of CR transport under different environmental conditions? 
\end{description}
We will conclude with some final remarks and an outlook on perspectives on the coming years in Section~\ref{sec:conclusions}.

This review will not cover the important and closely related aspects of CR origin and particle acceleration, treated in a companion paper in this volume \cite{DeOna:2021}, nor the propagation of CRs in the heliosphere \citep[for a review see][]{potgieter2013}.

Before entering the main matter, we define here some terminology that will be used in the paper.
\begin{description}
\item[Interstellar/diffuse emission:] we will refer to gamma-ray emission produced by CR interactions with interstellar matter and fields as interstellar emission. Conversely, we will use the term diffuse emission to refer to all emission that cannot be associated with a localized object (e.g., a pulsar, a binary system etc.) that is individually detected. Based on this definition, diffuse emission will comprise of interstellar emission plus collective emission from populations of sources not detected individually (due. e.g., to instrumental sensitivity limitations). 
\item[Large-scale galactic CR population:] we will use this term to refer to the CR population in a galaxy on spatial scales much larger than those where an individual CR source or sink (or localized groups thereof) can influence significantly the CR properties. The fact that this definition is useful in practice is based on observations of the Milky Way and local-group galaxies that will be discussed later in the paper. Conversely, we will avoid the term ``CR sea'', sometimes used in the literature with an ambiguous meaning that can refer to the CR population around the Earth, the large-scale galactic CR population (according to our definition), or the large-scale galactic CR population with an implicit assumption that it is uniform within (or even beyond) the galaxy. 
\end{description}

\section{The toolbox to study interstellar gamma-ray emission}\label{sec:obsmod}

\subsection{The progress of observational techniques in gamma rays}\label{sec:obsgamma}
Historically the most important facilities for the observations of interstellar gamma-ray emission have been space-based pair-tracking telescopes that cover the energy range from a few tens of MeV to tens of GeV and beyond. This is due to a combination of instrumental and intrinsic characteristics. Pair-tracking telescopes have a large field of view and a lower background than other gamma-ray detectors. Therefore, they are ideally suited to study diffuse emission, and for a long time they have been unrivalled in terms of sensitivity in the gamma-ray domain. Furthermore, in the GeV energy range we find the peak of energy output from CR interactions in the ISM, interstellar emission prevails over discrete sources, and it is dominated by hadronic emission correlated with interstellar gas (characterized by a well-defined morphology known from observations at other wavelengths, and, therefore, easier to separate from other emission components). The energy range covered by these instruments is often referred to as high-energy (HE) gamma rays. 

In the past decade advances in the HE range have been driven by the \textit{Fermi} LAT \cite{atwood2009}. Thanks to the use of Silicon tracking devices the LAT has reached in the GeV domain an unprecedented sensitivity, field of view (2.4~sr), and angular resolution (better than $\sim 0.8\degr$ at energies $> 1$~GeV and better than $\sim 0.15\degr$ at energies $> 10$~GeV). The LAT has also extended the energy reach of this observing technique up to TeV owing to a combination of instrumental improvements, notably the use of a segmented anticoincidence detector for CR background rejection.

Gamma-ray observations at lower energies require the use of space-borne telescopes exploiting different detection techniques: coded masks in the energy range from hundreds of keV to MeV and Compton detectors at MeV energies. In this domain the state-of-the-art instruments date back to twenty or even thirty years ago with \textit{INTEGRAL}~SPI\footnote{For \textit{INTEGRAL} legacy results see also \cite{integral}.} \cite{vedrenne1998} and COMPTEL \cite{schoenfelder1993}. Their performance cannot compete with the level reached by the LAT in the GeV domain \citep[e.g.,][]{knoedlseder2016}. New missions have been proposed to improve observational capabilities in the MeV-GeV energy range thanks to Silicon detectors that can carry out at the same time Compton and high-angular-resolution pair-tracking measurements, most notably ASTROGAM and AMEGO \cite{deangelis2017,mcenery2019,astrogam2021}. Alternatively, GECCO is a concept of combined dual mode telescope that can improve measurements in the sub-MeV to MeV energy range thanks to an innovative imaging calorimeter as a standalone Compton detector and, at the same time, as a focal-plane detector for a coded aperture mask  \cite{sasaki21}.

The limited size of space-borne instruments makes measurements at energies beyond several tens of GeV more and more difficult. Therefore, at higher energies ground-based instruments are used. Their energy range is often referred to as very-high-energy (VHE) gamma rays, or ultra-high-energy (UHE) gamma rays beyond 100~TeV. Observational techniques in this energy range are covered in detail in a companion paper in this volume \cite{Mirzoyan:2021}. Below we will summarize the most important aspects with emphasis on the study of interstellar emission.

Ground-based Imaging Air Cherenkov Telescope (IACT) arrays have proven to be a very effective way to study gamma rays above a few tens of GeV \citep[e.g.,][]{denaurois2015}. IACTs have fields of view of a few degrees and a high level of background due to CRs misclassified as gamma rays, and an angular resolution of several to a few arcminutes. This technique has reached its maturity with the current generation of arrays comprising 2 to 5 IACTs, namely H.E.S.S., MAGIC, and VERITAS. Among them, H.E.S.S., which is located in the Southern hemisphere and therefore has access to the inner part of the Milky Way,  has engaged a systematic survey of the Galactic plane and has achieved the detection of diffuse emission that is likely to be, at least in part, of interstellar nature \cite{aharonian2006,abramowski2014}. In this energy range, however, discrete sources prevail, thus we expect a sizeable fraction of diffuse emission to be due to unresolved sources not yet detected individually with the current sensitivity limitations (Section~\ref{sec:modda}).

The field of IACTs is going to be revolutionised in the next few years by the advent of the Cherenkov Telescope Array (CTA) \cite{actis2011,Acharya:2019}. CTA will feature more than one hundred IACTs of different sizes located on two sites in the Northern and Southern hemispheres, thus it will be able to observe the entire sky. It will cover the energy range from a few tens of GeV to $> 300$~TeV with a sensitivity one order of magnitude better than current IACTs, a field of view reaching $10\degr$, reduced CR background, and an angular resolution of a few arcmin.

A complementary observing technique for TeV gamma rays consists in ground-based shower particle detectors. Milagro has pioneered the use of water Cherenkov detectors \cite{abdo2012}, currently exploited by HAWC, which provides the best sensitivity among all existing instruments at energies $> 10$~TeV \cite{abeysekara2017}. Alternative approaches are the use of scintillator detectors adopted by the Tibet Air Shower Array \cite{amenomori1999}, or of resistive-plate counters adopted by ARGO-YBJ \cite{aielli2006}. These instruments have a large field of view, corresponding to the entire sky not occulted by the Earth, and a high duty cycle (contrarily to IACTs that operate only at night). Conversely, their angular resolution  is not as good as for IACTs. For example, for HAWC the angular resolution varies between $1\degr$ and $0.2\degr$ \cite{abeysekara2017}.

The LHAASO observatory, still under construction, combines shower particle detectors and IACTs. The expected steady-source sensitivity will be superior to that of CTA above a few tens of TeV \cite{cao2019}. All of the ground-based shower particle detectors mentioned so far were or are located in the Northern hemisphere. A new project has been proposed to install a water Cherenkov shower particle detector system in the Southern hemisphere, SWGO, which will then be able to observe the inner Milky Way and the Magellanic Clouds \cite{albert2019}.

\begin{figure}[p]
\widefigure
\includegraphics[width=0.75\textwidth]{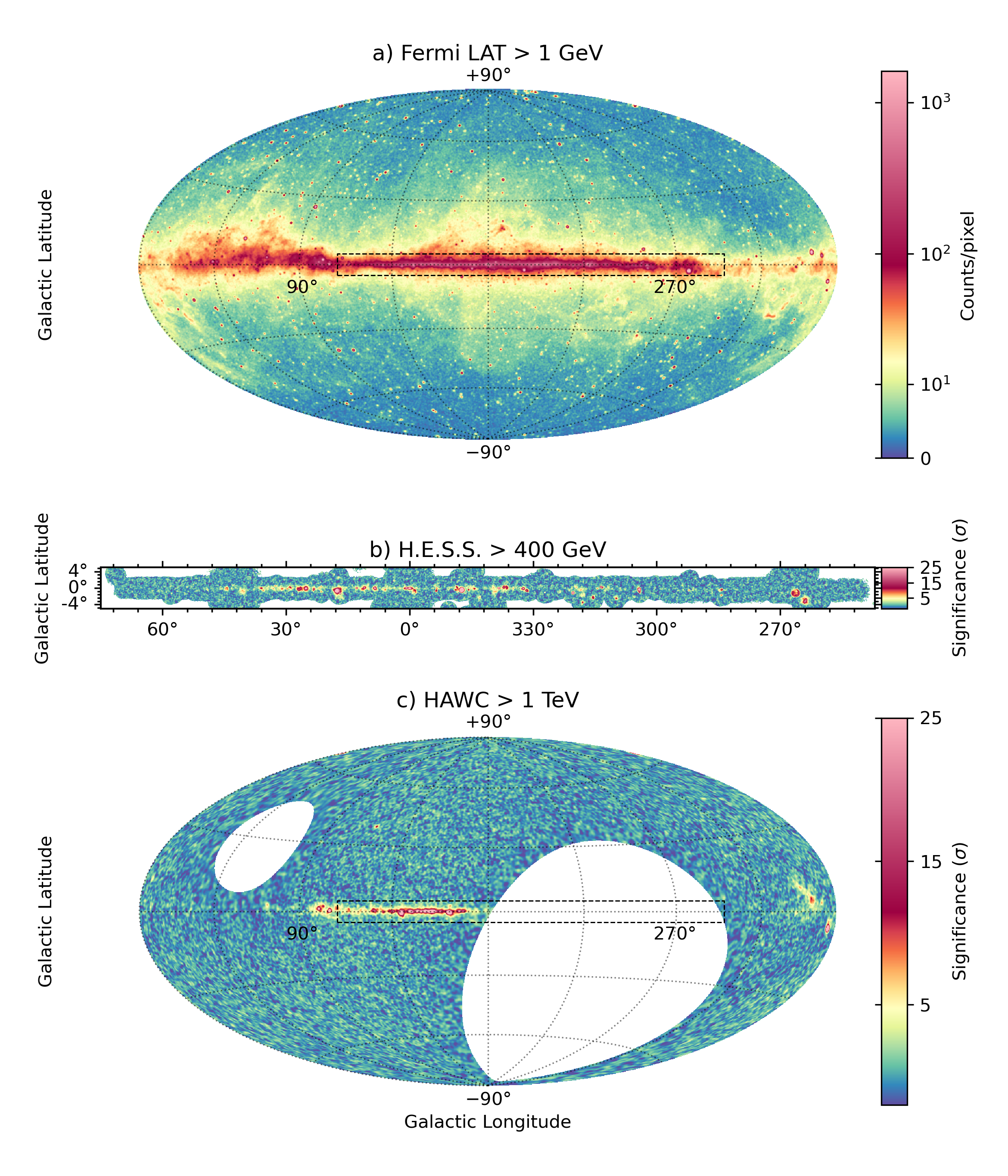}
\caption{\label{fig:skymaps}Images of the Milky Way from different instruments. (\textbf{a}) \F-LAT, 12 years of P8R3 data, energies $> 1$~GeV, \textit{Source}/PSF3 event class/type, zenith angles $<100\degr$, smoothed with a Gaussian kernel of $\sigma = 0.25\degr$. (\textbf{b}) H.E.S.S. Galactic plane survey, map described in \cite{hgps} ($0.2\degr$ correlation radius). (\textbf{c}) HAWC survey, map construction described in \cite{3hwc} (test source with  $0.5\degr$ extension and a power-law spectral index of $2.5$). The footprint of map (\textbf{b}) is overlaid to the all-sky maps in (\textbf{a}) and (\textbf{c}) as a dashed rectangle. Map (\textbf{a}) displays observed counts not corrected for residual CR background, while maps (\textbf{b}) and (\textbf{c}) are given in units of significance of gamma-ray emission above the residual CR background. For maps (\textbf{b}) and (\textbf{c}) the actual energy threshold varies across the map and the figure provides a representative value.}
\end{figure}

To illustrate the status of observations Figure~\ref{fig:skymaps} shows some recent maps of the Milky Way from different instruments.
The all-sky observing capabilities of the \F-LAT make it an invaluable source of information to study the entire range of manifestations of CR propagation and interactions. We note that features correlated with interstellar structures in the Milky Way are clearly visible in the map in Figure~\ref{fig:skymaps}(\textbf{a}) even though no background subtraction has been applied beyond event-wise selection of candidate photons.

\end{paracol}
\begin{paracol}{2}
\switchcolumn 

\F-LAT observations are complemented at higher energies by observations with IACTs, notably H.E.S.S., that for the moment can cover much more limited regions of the sky albeit with greater detail thanks to their superior angular resolution. Another limitation of IACTs is related to the presence of a much larger background due to CR interactions with the Earth's atmosphere. The traditional analysis techniques employed to deal with residual CR background are based on ``Off'' regions, most often chosen within the same field of view of the position of interest. The map  in Figure~\ref{fig:skymaps}(\textbf{b}) is constructed using ``Off'' regions in the shape of rings centered at the same position as the region of interest, or ``On'' region, that is a circle of fixed radius, sometimes referred to as correlation radius \citep[see ][for an illustration, in Figure~4, as well as for a review of the traditional background estimation techniques]{berge2007}. Due to the limited field of view of IACTs this results in a lesser sensitivity to extended emission \citep[see, e.g., the discussion in][]{jardin-blicq2019}. Alternative background estimation methods, either data-driven or simulation-based, are sought \cite{vovk2018,knoedlseder2019,mohrmann2019,holler2020}. The large times necessary to map large portions of the sky with IACTs and lesser sensitivity to extended emission makes their contributions to the study of large-scale interstellar emission less rich than those from wide-field of view instruments.
 
Shower particle detectors kick in at even higher energies offering a wide-field view of the sky as illustrated in the HAWC skymap shown in Figure~\ref{fig:skymaps}(\textbf{c}). In spite of a large CR background, the large field of view of these detectors makes it possible to routinely employ a data-driven background estimation method known as the ``direct integration'' technique \cite{atkins2003}, which exploits the facts that the CR background is stable in time and varies smoothly as a function of conditions in the atmosphere and the detector (e.g., trigger rate). Background estimation is performed independently in large declination bands, with an accuracy of order $10^{-4}$ limited by anisotropies in the primary CR arrival directions \cite{atkins2003}. This technique is therefore well suited for the study of large-scale emission.

\subsection{Data complementary to gamma rays: recent step forwards}\label{sec:otherobs}
The study of interstellar gamma-ray emission is deeply intertwined with other multi-messenger/multi-wavelength measurements and observations, that we can group in five broad categories:
\begin{itemize}
\item direct CR measurements;
\item the study of other ISM constituents: matter, radiation fields, magnetic fields;
\item other indirect CR tracers, that is, synchrotron emission, neutrinos, ion and nuclear lines;
\item the census of energetic objects such as massive stars, SNRs, pulsars and their wind nebulae;
\item hadronic interactions cross-sections.
\end{itemize}
In this section we will very briefly review these five domains and highlight some recent results, with emphasis on aspects of particular importance for the subjects covered in the review.

{\bfseries Direct CR measurements} assess the spectra, composition, and  arrival direction anisotropies of charged particles around the Earth. For sub-knee CRs this is prevalently achieved using satellite- or balloon-borne particle detectors, although ground-based instruments studying the byproducts of CR interactions in the atmosphere can also explore the energy range around the knee. 
The past few years have been marked by the high-precision measurements of CR spectra performed by AMS-02 on the International Space Station of a wide range of species, including: light nuclei \cite{aguilar2015H,aguilar2015He}, heavier nuclear species \cite{AguilarIron,AguilarFluorine,AguilarNeon}, electrons \cite{aguilar2014} and secondary species produced by CR interactions in the ISM \cite{aguilar2018}. They are complemented by new results on the abundances of heavy nuclei, e.g., \cite{murphy2016}.
Linking these measurements with gamma-ray observations is complicated by the fact that charged particles near the Earth below rigidities of $\sim$100~GV are affected by the solar wind, which modulates their spectra as a function of the solar cycle phase. This limitation has been overcome only recently for rigidities below $\sim$1~GV thanks to measurements of light CR species in interstellar space with the Voyager~1 probe. \cite{cummings2016}.
On the higher-energy portion of the spectrum, the TeV domain has also witnessed a significant advance, thanks to the measurements of several different balloon/satellite/ground experiments, both for hadronic species (e.g., from DAMPE \cite{An:2019wcw}, ATIC~\cite{Panov:2011ak}, and NUCLEON~\cite{Atkin:2018wsp}) and for leptons (e.g., from H.E.S.S.~\citep{Aharonian:2009ah,kerszberg_ICRC}, CALET~\citep{Adriani:2018ktz}, and DAMPE~\citep{Ambrosi:2017wek}).

{\bfseries Interstellar matter} constitutes a target for the production of gamma-ray emission via nucleon-nucleon inelastic collisions and electron Bremsstrahlung. The gamma-ray yield is proportional to the mass which resides predominantly in gas, the most important contributor being hydrogen, either in the atomic, molecular, or ionized form. {\it Atomic hydrogen} (\hi) is widely distributed in galaxies, and can be traced thanks to the 21-cm hyperfine transition line. The velocity-integrated brightness temperature of the 21-cm line is directly proportional to the gas column density in the optically-thin limit. Often we need to account for \hi opacity, which is typically done under the approximation of a uniform spin temperature. Recent years brought remarkable advances in the observations both for the Milky Way at large \cite{hi4pi,galfa}, and for specific Galactic or extragalactic regions. {\it Molecular hydrogen}, \hd, mostly concentrated in cold clouds, is difficult to observe directly and is most often traced indirectly using molecular lines of other species. The mm rotational lines of the second most abundant interstellar molecule, CO, with its different isotopes, have been a major tool in gamma-ray astronomy. While for Milky Way on large scales we still rely on the survey by \citet{dame2001}, high-resolution surveys of specific regions or external galaxies become more increasingly available, e.g., \cite{furukawa2014,mopradr3}.
It is empirically established that molecular hydrogen column densities,  \nhd, are approximately proportional to the velocity-integrated brightness temperature of the $^{12}$CO $J=1\rightarrow 0$ line, \wco, via the \xco$=$\nhd/\wco factor, for which, however, variations are both observed and expected \cite{bolatto2013} (see also the discussion in Section~\ref{sec:locemiss}). {\it Ionized hydrogen}, \hii, is present in regions around star-forming regions and in a kpc-wide layer around the Galactic disk. The $\mathrm{H}\scriptstyle{\alpha}$ recombination line in the visible band is heavily absorbed in the ISM. Alternative tracers are provided, under different kinds of hypotheses and approximations, by microwave free-free emission \cite{sodroski1997}, pulsar dispersion measurements \cite{ne2001}, and radio recombination lines \cite{alves2015}. Owing to Doppler shift from the Galactic rotation, lines of all kinds can be used to separate different structures along the line of sight, and approximately locate them.

An alternative approach to trace interstellar matter relies on {\it dust}. Dust grains make a tiny fraction of the mass in the ISM, but they produce bright thermal emission in infrared and are responsible for stellar extinction in the near-infrared to visible domain. They are thought to be well mixed with gas, therefore their emission/extinction can be used as a tracer of total ISM masses. Recent observational developments include the improved mapping of thermal dust emission thanks to the \textit{Planck} satellite \cite{planckcompsep} and strong advances in 3D dust mapping based on stellar extinction measurements combined with stellar population synthesis models, e.g., \cite{green2019,lallement2019}.  As for \xco, variations in the ratios between dust observables and matter column densities are both observed and expected, e.g., \cite{remy2018}. The combination of gamma-ray and dust observations (both tracers of the total masses in the ISM) with the \hi and CO lines has demonstrated that the aforementioned lines fail to trace the totality of the neutral interstellar medium. The excess with respect to the \hi- and CO-bright gas is known as dark gas or {\it dark neutral medium} (DNM). It is predominantly located at the interface between the molecular-dominated and atomic-dominated parts of interstellar clouds, and it is likely made of a combination of optically-thick \hi and CO-poor \hd in the outer layers of the molecular regions less shielded from UV photo-dissociation \citep[for a recent review see][]{grenier2015}. The existence of the DNM is confirmed by alternate molecular tracers, e.g., \cite{balashev2017,engelke2019} and emission from ionized carbon, e.g., \cite{bigiel2020}, and also supported by numerical simulations, e.g., \cite{smith2014}.

{\bfseries Interstellar radiation fields (ISRFs)} constitute a target for the production of gamma-ray emission via inverse-Compton scattering by CR electrons and positrons. They include the cosmic microwave background, thermal emission from dust grains heated by stellar radiation from sub-mm to infrared, and radiation from stars from near infrared to UV. Radiative transfer techniques can be used to link the ISRFs to the measured spectral energy distributions and other observational constraints on the spatial distribution of stars and interstellar dust. Recent years have seen significant advances in this field both on the observational front, notably with the improved measurements of thermal dust emission thanks to the \textit{Planck} satellite \cite{planckcompsep}, and on the modelling front for the Milky Way \cite{vernetto2016,popescu2017,porter2017} as well as for nearby external galaxies, e.g., \cite{popescu2011,paradis2011}. The importance for gamma-ray observations was discussed recently by \citet{niederwanger2019}.

{\bfseries Magnetic fields} are relevant to CRs both as a target for synchrotron energy losses/radiation and as the agent of diffusion (see Section~\ref{sec:crtheory} for a discussion on the latter). We can separate interstellar magnetic fields into a large-scale regular and a turbulent component. In external galaxies they are known to follow a spiral structure similar to that of interstellar matter and stars \citep[for a recent review see][]{beck2015}. The origin of the regular field  is still debated. It is constrained through rotation of polarized emission from background sources, e.g., pulsars, polarized synchrotron emission, and polarized dust emission. Many recent works have used observational constraints to model the large-scale Galactic magnetic field, e.g., \cite{sun2010,fauvet2012,janssonfarrar2012,jaffe2013,orlando2013}, and have been revisited on the light of \textit{Planck} results \cite{planckregularb}. Magnetic turbulence is thought to be driven by supernova explosions, stellar winds and outflows, shocks and instability induced by galactic rotation, and shear instabilities and baroclinic effects in the ISM. It is related to interstellar turbulence in velocity, matter density, and free-electron density. It is therefore constrained observationally using high-resolution spectroscopy of  interstellar lines, high-resolution imaging of interstellar matter, intensity and polarization of synchrotron and dust thermal emission,  dispersion of pulsar signals, interstellar scintillation, and rotation of polarized emission from background sources \citep[for a recent review see, e.g., ][]{ferriere2020}. The combination of these observations reveals an overall power-law spectrum as a function of wavenumber with Kolmogorov slope over spatial scales from thousands of km to a few pc \cite{chepurnov2010}.

{\bfseries Indirect CR tracers} other than gamma-ray continuum emission include the already mentioned {\it synchrotron emission}, observed from radio to microwaves. The microwave sky was recently studied in unprecedented detail by the \textit{Planck} satellite. Studies of synchrotron emission are used to reconstruct the broadband spectrum of CR electrons and to inform the interpretation of observations of IC emission in gamma rays \cite{orlando2018,orlando2019sync}. Only a few years ago the first detection of {\it astrophysical high-energy neutrinos} with IceCube \cite{icecube2013} has opened a new window that may enable us to have a complementary tracer of CR nuclei in galaxies, but for the moment only upper limits to a Galactic neutrino signal exist combining data from IceCube and Antares \cite{icecubeantares2018}. {\it Molecular line emission} driven by CR ionization, which yields, e.g., $\mathrm{H}_3^+$, OH$^+$, $\mathrm{H}_2\mathrm{O}^+$, $\mathrm{H}_3\mathrm{O}^+$, is observed in infrared and mm waves and provides information on the low energy part of the CR spectrum \citep[for a recent review see ][]{grenier2015}. Recent calculations show that the observed molecular ion line emission suggests an average ionization rate a factor of 10 larger than what is expected from directly measured CR spectra (including results from Voyager~1) \cite{neufeld2017,phan2018}. This may point to the existence of an additional CR component emerging at low energies different from those observed directly or through gamma rays, although it seems more likely that ionization sources different from CRs may play a role more prominent than previously thought \cite{recchia2019} Furthermore, we note that the methodology used to infer the ionization rate from the data is very sensitive to the composition of the ISM, e.g., to the presence of polycyclic aromatic hydrocarbons \cite{shaw2021}. An alternative way to study the CR nuclei population in remote locations below the pion production threshold (kinetic energies of $\sim$300 MeV/nucleon) would be to observe {\it nuclear de-excitation lines} in the $0.1-10$~MeV range induced by CR collisions with interstellar matter \cite{tatischeff2004,liu2021} thanks to a future MeV telescope \cite{orlando2019}.

{\bfseries Energetic objects} play a twofold role: they are potential CR accelerators and they inject energy into the ISM under other forms, e.g., radiation and magnetic turbulence. Knowledge of their census and its recent history is therefore essential to model interstellar gamma-ray emission and interpret the gamma-ray observations. Different challenges are to be faced to determine the distribution in space and time of energetic object at galactic scales against observational uncertainties and biases, or to establish a detailed picture of individual remarkable regions. Among the different classes of energetic objects, massive stars are interesting at the same time for themselves, and as the progenitors of all other relevant classes such as SNRs and pulsars and their wind nebulae. Our view of stellar populations in the Milky Way is in a transformative phase thanks to the data collected by the \textit{Gaia} satellite in the visible/near-infrared band, which provide precision measurements of positions, parallaxes, and proper motions of over 1.4 billions stars within $> 4$~kpc around the Solar system \cite{gaiaedr3}. This makes it possible to paint a portrait of the history of stellar clusters in the disc of the Milky Way over the past billion years \cite{cantat-gaudin2020}, which complements information from observations in near-infrared, e.g., \cite{kharchenko2013}, or at lower frequencies, e.g., \cite{reid2019}. At the cluster or star-forming region level, this enables us to go beyond simple models of coeval and colocated star formation, and embrace more realistic descriptions of its spatial \citep{Berlanas:2019} and temporal distributions \citep{Beasor:2021}. For SNRs, pulsars and pulsar wind nebulae (PWNe) the most important wavebands for the observations are radio, X rays, and gamma rays. For the first two bands the coming decade is expected to be marked by the results from the Square Kilometer Array \cite{ska2015} and its precursors and pathfinders\footnote{\url{https://www.skatelescope.org/precursors-pathfinders-design-studies/}}, and the eROSITA space telescope \cite{eROSITA}. The role of gamma-ray observations in understanding particle acceleration in energetic objects is treated in a companion paper in this volume \cite{DeOna:2021}.

{\bfseries Hadronic cross sections} for gamma-ray production are an essential ingredient to model and interpret gamma-ray observations. While leptonic cross sections in principle can be calculated exactly, the modelling of hadronic interactions heavily relies on experimental constraints. Accelerator data provide information with a somewhat limited coverage in terms of energies, angular distribution, and interacting species (mostly p-p), that are then used to model the cross sections resulting in non-negligible uncertainties\citep[for a recent review see][]{dermer2013}. Accelerator data in the crucial energy range above the pion production threshold and around the $\Delta(1238)$ isobar resonance, and up to center-of-mass energies of $10^3$~TeV mostly date back to the '50s-'80s and have been compiled by \citet{lock1970}, \citet{stecker1973}, and \citet{dermer1986}. The energy coverage was recently extended up to $10^8$~TeV in the center-of-mass frame thanks to the LHCf experiment \cite{adriani2011,sato2012}, and improved also at hundreds GeV energies thanks to the NA61/SHINE experiment  \cite{na612017}. Recent cross-section derivations exploiting these data include \citet{kamae2006},  \citet{kelner2006},  \citet{kachelriess2012}, \citet{kafexhiu2014}, \citet{mazziotta2016}, \citet{kachelriess2019}, and, with focus on the contributions of species heavier than protons, \citet{mori2009}, and \citet{kachelriess2014}.

\subsection{A glimpse at the basics of cosmic-ray transport}\label{sec:crtheory}

Gamma-ray emission from the galactic ISM is intimately associated with the physical problem of CR acceleration and transport. The problem of CR acceleration is not covered in this review.  We just recall that the SNR paradigm is widely considered as the reference guideline, although other classes of sources powered by a variety of mechanisms have been proposed as well (e.g. OB associations, X-ray binaries, and pulsar wind nebulae for leptonic CRs). Within the SNR scenario, the theory of  diffusive shock acceleration \cite{blandford1978,bell1978,axford1977,krymskii1977} and its nonlinear extension \cite{Blasi:2002} describe the process o acceleration of cosmic particles that are diffusing around the forward shock in an SNR, and are able to reproduce the correct CR energetics and overall many of the CR observables. We refer to \citet{Blasi:2013} for an extensive review on the origin of CRs and the SNR paradigm, and also to  \citet{DeOna:2021} in this volume for the role of gamma-ray observations in this context. 

In this Section we focus instead on the basics of the problem of galactic CR transport. Let us start by mentioning that a large body of evidence demonstrates that high-energy charged CRs are confined in the Milky Way for a timescale that is much longer than the ballistic crossing time. In fact, the analysis of the properties of the CR fluxes that reach Earth outlines the following key features:
\begin{itemize}
\item isotropy at the level of $\sim 10^{-3}$ in the arrival directions for the entire energy range covered by the experiments that are sensitive to this observable, in particular in the TeV - PeV range \cite{anisoKASCADE,anisoMILAGRO,anisoEASTOP,anisoTIBET,anisoIceTop,anisoIceCube}; this suggests that particles have suffered multiple deflections in their journey from the acceleration sites to Earth;
\item a significantly larger abundance of some light species, namely Lithium, Beryllium, and Boron, with respect to the solar system abundances; this piece of evidence in particular is naturally interpreted as the smoking gun of the cumulative interactions between the primary species injected by the accelerators (protons and heavy nuclei) and interstellar gas; the total column density that primary CRs have to cross to produce the observed amount of light species is as large as a few g/cm$^2$: such a large value of the {\it grammage} strongly suggests once again that the CRs that produce the secondary species have crossed the Galactic disk multiple times over time scales exceeding one Myr. 
\end{itemize}

The picture is corroborated by the ubiquitous observation of magnetic turbulence in the interstellar environment that we briefly mentioned in the previous section.
The multiple, random interactions of charged CRs with these perturbations naturally provide a mechanism to explain why the motion of these particles should be described as a diffusive process.
These considerations, together with an increasing amount of data in different channels discussed in the previous section, corroborate the standard paradigm that seems to capture the most relevant aspects of  Galactic cosmic-ray physics, as recently extensively reviewed in \cite{Gabici:2019jvz}, 
in which CRs are diffusively confined within a kpc-sized, magnetized and turbulent Galactic halo. 

The simplest way to describe this phenomenon, widely used in the past literature, is provided by so-called {\it leaky-box models} \cite{Cowsik1967}. In this framework, the galaxy is modelled as a cavity with almost perfectly reflecting ``walls''. The cosmic particles are allowed to move freely within this environment. The physics of their propagation and escape is all embedded in the energy-dependent parameter $\tau_{esc}$, i.e., the mean residence time. Thus, the probability of particle escape per unit time is equal to $\tau_{esc}^{-1}$. The model is described by the following equation for the particle density $N$:
\begin{equation}
\frac{\partial N}{\partial t} (E) \,=\, \frac{N}{\tau_{esc}(E)} + Q(E),
\end{equation}
where $Q(E)$ is the source function. 

Recently, a description in terms of diffusion has become prevalent. We want to emphasize that it is very challenging to obtain a general expression of the diffusion tensor from first principles. 
A widely used guideline in this context is the {\it quasi-linear theory of pitch-angle scattering} onto magnetic fluctuations presented in the pioneering papers of \cite{Jokipii1966,Jokipii1968}.  
The rationale of the theory is to consider the interaction of a charged particle of momentum $\vec{p} = \gamma m \vec{v}$ with magnetic inhomogeneities $\delta \vec{B}$. 
The key assumptions behind this theoretical framework are the following:
\begin{itemize}
\item the magnetic inhomogenities are Alfv\'enic; they are isotropic and their energy density is characterized by a power-law energy spectrum $P(k)$ as a function of wavenumber $k$;
\item the inhomogenities are small, at the scale of interest, with respect to the coherent large-scale regular magnetic field $B_0$.
\end{itemize}

The key result of this approach is that the particles mainly diffuse along the regular magnetic field. 
The process is resonant, i.e., the Alfv\'en wavepackets that contribute to the process have a wavelength comparable to the gyroradius of the particle. It is useful to notice that the length scales associated to the energy range usually covered by current CR observations are typically very small compared to the size of a galaxy, and to the scale of injection of turbulence (10-100 pc). For instance, GeV particles resonate with fluctuations with wavelength of the order of few AU.

The resulting scattering rate can be written as \cite{1964ocr..book.....G,Ginzburg:1990sk,Blasi:2013}:
\begin{equation}\nonumber
\nu = \frac{\pi}{4} \, \frac{k_\mathrm{res} P(k_\mathrm{res})}{B_0^2/(8 \pi)} \Omega_g
\end{equation}
where $\Omega_g = q B_0 / (\gamma m c)$ is the  gyration frequency and the resonant wavenumber is $k_\mathrm{res} = \Omega_g / v_\parallel$  ($v_\parallel$ is the velocity component along the coherent magnetic field $B_0$). 

Starting from this expression, it is possible to obtain a (parallel) spatial diffusion coefficient of this form:
\begin{equation}\nonumber
D(p) = \frac{v^2}{3 \Omega_g} \frac{B_0^2/(8 \pi)}{k_\mathrm{res} P(k_\mathrm{res})}.
\end{equation}
It is useful to recast this expression into:
\begin{equation}\nonumber
D(p) = \frac{1}{3}  \frac{r_L v}{\mathcal{F}(k_\mathrm{res})}
\end{equation}
where $r_L = p_{\perp}/qB_0 $ is the Larmor radius of the particle and we have defined 
\begin{equation}\nonumber
\mathcal{F}(k) \equiv \frac{k P(k)}{B_0^2/(8 \pi)}.
\end{equation}
This expression shows that a larger power in magnetic fluctuations at a certain scale is associated to a lower diffusion coefficient for the resonating particles, hence a more effective confinement. The dependence on the Larmor radius, both direct and indirect via the resonant wavenumber $k_\mathrm{res}$, and the empirical power-law dependence on wavenumber of the magnetic turbulence spectrum observed at large scales drive a dependence of the diffusion coefficient on particle rigidity $R$. Standard implementations for the Milky Way feature a diffusion coefficient $D(R) = \mathcal{O}(10^{27})\, \beta\,  (R/1\,\mathrm{GV})^{1/3}$ cm$^2$ s$^{-1}$ in reasonable agreement with a reference estimate of the random field at the injection scale and extrapolation down to the resonant scale.
 
We emphasize that the theory is typically built on an isotropic picture of turbulence. However, the resulting process is highly anisotropic. We will elaborate more on these key concepts in the next Section.

Diffusive confinement is certainly a key feature characterizing CR propagation. 
However, all CR species interact in many different ways with the different components of the ISM, and a variety of other processes occur during their random walk across the parent galaxy.
Let us briefly recap the most relevant ones.
\begin{itemize}
\item {\bfseries Reacceleration: } This process is intimately connected to spatial diffusion. In fact, the random walk in space is expected to be accompanied by a diffusion in momentum space, since the scattering centers (namely, magnetic fluctuations) are not static. They are instead in random motion themselves, with characteristic velocities of the order of the Alfv\'en speed. The importance of this process hence depends on the large-scale average of this quantity over the galaxy, and has been the subject of a long debate. We refer to \cite{Drury:2016ubm} for a critical look at this issue in the case of the Milky Way, in connection with the total energy budget available in the Galaxy.
\item {\bfseries Advection: } This is a rigidity-independent process that can significantly contribute to the vertical escape of CRs and is associated to the existence of so-called galactic winds.  This phenomenon consists in a powerful outflow that may extend for hundred of parsecs, possibly more relevant in the inner part of galaxies, and induce a relevant mass loss, possibly comparable to the mass formed in stars. Possible mechanisms to create galactic winds are currently under debate. Winds may be powered by stellar winds and supernova explosions, and a non-linear interplay with CRs may also be at work:  cosmic rays (CRs) escaping from the galaxy can effectively push on the ISM and eventually trigger the wind itself. 
\item {\bfseries Energy losses: } Different types of interactions transfer energy from the CR population to the ISM. In particular, hadrons loose energy due to ionization, Coulomb interactions with interstellar matter, and production of secondary particles in nucleon-nucleon inelastic collisions, most notably pions (these processes being overall more effective at low energy, in particular in the sub-GeV domain). On the other hand, leptons suffer strong losses mostly due to IC scattering onto low-energy photons, synchrotron emission (with a rate that increases with increasing energy, following a $\propto E^2$ scaling), and Bremsstrahlung. We refer for instance to \cite{Evoli2017I} for a compilation of the relevant formulae associated to these processes.
\item {\bfseries Spallation: } A complex network of nuclear reactions and decays transform heavier CR nuclei into lighter species as a consequences of inelastic interactions with matter in the ISM. A combination of semi-empirical parametrizations and rescaling procedures to nuclear data (mostly available in the GeV domain) is typically adopted to model these phenomena. See for instance \cite{Silberberg1990,Strong:2001fu,Mashnik:1998sm,Moskalenko:2001qm,Moskalenko:2003kp,Evoli2017II} and references therein for the modelling of the hadronic nuclear network. 
\end{itemize}

Remarkably, a joint description of the most relevant phenomena mentioned above  is possible in the form of a relatively compact {\it transport equation} that can be solved for each CR species of interest.
This general reacceleration-diffusion-loss equation is usually written as follows:~\citep{1964ocr..book.....G,Ginzburg:1990sk}.
\begin{equation}\label{eq:prop}
\begin{array}{r}
- \bm{\nabla} \cdot ( D \bm{\nabla} N_i \,+\, \bm{v}_w N_i) + \frac{\partial}{\partial p} \left[ p^2 D_{pp} \, \frac{\partial}{\partial p} \left( \frac{N_i}{p^2} \right) \right] - \frac{\partial}{\partial p} \left[ \dot{p} N_i - \frac{p}{3} \left(\bm{\nabla} \cdot \bm{v}_w \right) N_i \right] = \\
Q + \sum_{i<j} \left( c \beta n_{\rm gas} \, \sigma_{j \rightarrow i} + \frac{1}{\gamma \tau_{j \rightarrow i}} \right) N_j - \left( c \beta n_{\rm gas} \, \sigma_i + \frac{1}{\gamma \tau_i} \right) N_i 
\end{array}
\end{equation}

In this equation: $p$ is the particle momentum; $N_i$ is the CR density for species $i$; $D$ is the spatial diffusion tensor; $D_{pp}$ the diffusion coefficient in momentum space, associated to reacceleration; $\bm{v}_w$ the velocity associated to advection; $Q$ is the source term; $(\sigma_{j \rightarrow i}, \sigma_i)$ are the spallation cross sections associated, respectively, to the creation of the species $i$ from parent nucleus $j$, and to the destruction of the species $i$; $(\tau_{j \rightarrow i}, \tau_i)$ are the decay times for, respectively, the unstable species $j$, creating $i$, and for $i$, creating lighter nuclei. For a detailed discussion on each term, we refer to the technical papers cited above.

\subsection{Evolutions of modelling and data analysis techniques}\label{sec:modda}
Extracting properties of CRs from the gamma-ray data always involves some type of modelling. The modelling of CR propagation and interactions and the associated non-thermal emission can follow different avenues. 
We will review in this Section different methods and discuss the most remarkable achievements and open questions.

The {\bf template fitting} method is a widely-used technique aimed at modelling observations of interstellar gamma-ray emission. In its simpler form the key assumption is that CR densities vary mildly on the spatial scales characteristic of interstellar gas complexes. Therefore, gamma-ray emission associated with interstellar gas can be modelled as a linear combination of maps (templates) of gas column density, or a proxy thereof, split for different regions along the line of sight by using the Doppler shift information of interstellar lines. The original implementation of this method was using templates derived from the \hi 21-cm line to account for atomic gas and from the 2.6-mm CO line as a surrogate tracer of molecular gas \cite{lebrun1983}. The linear combination coefficients, known as gamma-ray emissivities, encode information on CRs. Notably, the \hi emissivity is nothing else that the gamma-ray emission rate per hydrogen atom, i.e., the convolution of the CR densities with the gamma-ray production cross sections. The gamma-ray analysis can be performed over several independent energy bins to reconstruct the underlying CR spectrum. In recent years the template fitting method has been extended to account for other forms of interstellar gas (dark neutral medium, ionized gas), and other components of interstellar emission, notably IC emission. For the latter templates need to be obtained using predictive models. More recently, the {\tt SkyFACT} tool \cite{Storm:2017arh} introduced the possibility to allow pixel-by-pixel variation within each template guided by a  penalized likelihood maximization by combining methods of image reconstruction and adaptive regression. Crucial uncertainties affecting the template fitting technique come from approximations in the construction of the templates, and cascade effects that stem from those in the component separation procedure. 

A specular technique consists in making assumptions about the spectra of different gamma-ray emission components, either based on theory or observations, and use the gamma-ray data to infer the morphology of the components. This technique is known as spectral component analysis or spectral template fitting \cite{malyshev2012,deboer2017}. This alternative incarnation of template fitting has been used less widely owing to the absence of sharp spectral features in typical gamma-ray spectra and because the $\gtrsim$5-10\% energy resolution of gamma-ray telescopes makes it less effective. We warn the reader that caution is needed in dealing with the energy-dependent point sprwad function (PSF) of the instruments when applying this technique.

Another completely {\bf data-driven} technique aimed at studying gamma-ray emission is represented by the D$^3$PO (Denoised, Deconvolved, and Decomposed) inference algorithm presented in \cite{Selig:2014qqa}. This method performs a Bayesian inference  without the use of  templates. Instead, it is designed to remove the shot noise, deconvolve the instrumental response, and finally provide estimates for the different flux components separately. This method is particularly suited in identifying and subtracting point sources from the data to study the remaining emission.

Let us now turn our attention to another widely adopted approach to gamma-ray modelling, which is the use of {\bf predictive models} connected to the physics of CR propagation in galaxies. The rationale of these methods is to compute the equilibrium CR distribution in the galaxy by solving the transport equation (Equation~\ref{eq:prop}) presented in Section 2.3. We have discussed how such equation captures the variety of physical processes shaping the transport of the cosmic particles from their production at the accelerator sites to their eventual escape from the large-scale diffusive halo.
Today we have at our disposal several public numerical codes, equipped with different numerical methods and astrophysical ingredients, aimed at solving that equation for all CR species in the Milky Way. The most important are (in chronological order): \texttt{GALPROP}\cite{Galprop1,Galprop2,Galprop3}, \texttt{DRAGON} \cite{Evoli:2008dv,Gaggero:2013rya,Evoli2017I,Evoli2017II}, \texttt{PICARD} \cite{Picard1,Picard2}. A semi-analytical approach is instead followed by the \texttt{USINE} project \cite{Usine}. 

The models based on this concept were remarkably successful in reproducing a variety of local CR data, and, for some of them, in modelling the non-thermal emission from the radio band all the way up to the GeV-TeV gamma-ray domain. In standard implementations the CR transport is typically described as isotropic, homogeneous diffusion characterized by a scalar diffusion coefficient with a power-law rigidity dependence. The amplitude and slope of the diffusion coefficient, together with a set of parameters associated to the astrophysical ingredients of the model (for instance, the \xco conversion factor between the CO emission intensity and the molecular gas column density) are typically fitted to a variety of data, including the accurate dataset of secondary/primary nuclei provided by AMS-02, and the gamma-ray maps provided by the \F-LAT.

However, a number of anomalies have been highlighted over the recent years both in CR and gamma-ray data: a break in the local proton, Helium, light and heavy nuclei spectra at $\sim$200 GV \cite{aguilar2015H,aguilar2015He,Aguilar:2018keu}, an excess of high-energy positrons first highlighted by PAMELA  \cite{Adriani:2008zr} and then confirmed by \F-LAT and AMS-02 (with better statistics) \cite{AMSpositrons}, a hint of an antiproton excess near 80 GeV \cite{Cuoco:2016eej,Cuoco:2019kuu}, and several observations in gamma rays that will be discussed in detail in Section~\ref{sec:largescale}.

Some of these anomalies posed a challenge to standard implementations of Galactic CR models, and have spurred a variety of new developments. 

\begin{itemize}

\item {\bf Three-dimensional modelling.} The transport equation is usually solved under the assumption of cylindrical symmetry. In this widely used setup, that is a distinctive feature of standard Galactic CR model implementations, most astrophysical ingredients that enter the problem and influence the different types of CR interactions (for instance, the distribution of CR accelerators, the magnetic field strength, the interstellar gas and low-energy photon distributions) are implemented in the form of (smoothly-varying) functions of the Galactocentric radius $R$ and the vertical coordinate $z$ (perpendicular to the Galactic plane).
A more realistic description of the interstellar medium, featuring a three-dimensional model for the spiral arm pattern of the Milky Way in the CR source term was first introduced in the context of numerical modelling of leptonic CR species in \cite{Gaggero:2013rya}, showing a relevant impact on the local electron spectrum. 
The consequences of such three-dimensional pattern on CR hadronic species and gamma-ray modelling was later discussed in \cite{Kissmann:2015kaa,Kissmann:2017ghg}. The authors of \cite{Porter:2017vaa} further investigated the phenomenological consequences of a spiral arm pattern in both the CR source distribution and the interstellar radiation field (see also \cite{orlando2019sync}). 

\item {\bf Inhomogeneous diffusion.} The transport equation is usually solved under the assumption that the diffusion coefficient is constant in both slope and normalization across the Galaxy. 
Several steps towards a inhomogeneous description of the problem have been proposed over the years. For instance, in \cite{Evoli:2012ha} the radial variation of the diffusion coefficient normalization was explored as a possible solution to the long-standing anomaly  usually identified as {\it gradient problem} (Section~\ref{sec:largescale}). 
Moreover, in \cite{Johannesson:2016rlh}, a complete scan of the parameter space for CR injection and propagation showed that each set of species is probing a very different interstellar medium, providing further evidence in favour of non-homogeneous transport. 


\item {\bf Anisotropy of CR transport.} The anisotropic nature of CR transport has been highlighted by theorists since the first pioneering studies about the quasi-linear theory of pitch-angle scattering onto magnetic fluctuations.  We have stressed in the previous section that the main feature of such theory is precisely a diffusive motion that is {\it strongly anisotropic} locally and goes predominantly along the magnetic field lines. 
However, since relevant fluctuations of magnetic field on scales close to the ones associated to the injection of turbulence may in principle lead to some isotropization of global CR diffusion in the Galaxy, the typical approach in the context of large-scale modelling of CR transport based on the transport equation has been to neglect any anisotropy in the diffusion part (see for instance the discussion in \cite{Strong:2007nh}).
Very few attempts to consider anisotropic transport aimed at reproducing local CR data by solving the large-scale diffusion equation exist in the literature. A relevant example is \cite{Cerri:2017joy} where a fully anisotropic diffusion equation is solved numerically in cylindrical symmetry.

\item {\bf CR self-confinement and non-linear effects.} We have mentioned that the magnetic irregularities (typically Alfv\'en waves) play a central role in the current description of CR confinement in the Galaxy. Those waves may be either part of a pre-existing turbulent cascade, or generated by the CRs themselves, if they stream faster than the Alfv\'en speed, via the process of {\it streaming instability} \cite{Wentzel:1974cp}. The latter option may give rise to non-linear effects in the CR transport, typically refereed to as {\it CR self-confinement}.
This phenomenon can dominate the transport at low energies, and can be responsible for a spectral feature at the energy where the scattering onto pre-existing turbulence start to dominate (see for instance \cite{Farmer:2003mz}). An interesting interpretation of the hardening in the CR spectra observed by the AMS collaboration was presented in \cite{Blasi:2012yr}. An even more refined description of CR transport in the Galaxy where the (coupled) CR diffusion equation and the equation for self-generated waves are solved numerically is presented in \cite{Evoli:2018nmb}. In this work, the size of the diffusive halo is derived from the model, as the result of a combination of wave self-generation and advection from the Galactic disc. These effects take an even more prominent role in the proximity of CR accelerators (Section~\ref{sec:youngcr}).


\item {\bf Anisotropy of the turbulent cascade. CR scattering onto magnetohydrodynamic turbulence modes.} The nature of the pre-existing magnetic fluctuations responsible for CR confinement in the high-energy range is an important matter of debate. 
According to the widely accepted model of turbulence by Goldreich and Shridar \cite{GS1994ApJ...432..612S,GS1995ApJ...438..763G}, the key feature of the Alfv\'enic cascade is its {\it anisotropic} nature\footnote{For the sake of clarity, we remark here that, in this case, the anisotropy of the turbulent cascade is the key point. This is a different concept with respect to the anistropy of the CR transport phenomenon. In fact, as seen in our recap on the quasi-linear theory, an isotropic cascade of turbulent fluctuations can yield a strongly anisotropic CR transport.}. Since most of the  power is transferred to scales perpendicular to a mean-magnetic-field direction, the Alfv\'enic waves may actually be highly inefficient in confining CRs, as discussed extensively in \cite{Chandran2000PhRvL..85.4656C,PhysRevLett.89.281102}, and pitch-angle scattering onto magnetosonic modes may play the dominant role. 
A non-linear theory of CR scattering onto magnetosonic modes was presented in \cite{Yan:2004aq,Yan:2007uc}. Very recently, after the seminal attempt presented in \cite{Evoli:2013lma}, the authors of \cite{Fornieri:2020wrr} provided the first comprehensive phenomenological study of such theory, and showed how local CR data above the AMS break can be reproduced by solving the aforementioned diffusion equation with the diffusion coefficients computed {\it ab initio} from the theory, under reasonable assumptions on the free parameters involved. Recent radio observations support the notion that magnetosonic modes, under some circumstance, may drive CR transport \cite{Zhang:2020}.


\item  The {\bf Monte Carlo approach.} The approach of solving the transport equation in a continuous setup where all the relevant terms are provided as smoothly varying function of the position provides a well-defined prediction of the expected {\it average} flux of cosmic rays. However, the stochastic nature of sources may play a relevant role in several cases, depending on the type of particle and on the energy range. For instance, high-energy leptons may be highly sensitive to this aspect, especially at energies at which the characteristic time and length scales associated to their momentum losses and spatial diffusion become comparable with the mean spatial and time distance between two different CR injection episodes. An important question is therefore to assess the expected {\it variance} of the CR flux, and a useful technique to attack this question is a Monte Carlo simulation. In this approach, many stochastic realizations are considered. In each of them,  a random set of acceleration events in considered, and the CR flux from each event is typically evaluated by means of an analytic formula and added up. Some relevant examples of works based on this technique are \cite{Ptuskin2006,Busching2005,BlasiAmato2012,Evoli2020}. The observables that are investigated are the fluctuations of the CR spectrum and normalization, anisotropy, chemical composition (especially around the knee).

\end{itemize}

To conclude, let us discuss briefly another broad class of modelling/data analysis methods that concerns the treatment of {\bfseries populations of unresolved sources} detected by instruments in the form of diffuse emission that needs to be disentangled from interstellar emission. A first approach to this challenge is to develop source population synthesis models constrained by bright sources already detected, and then use them to predict the unresolved component, e.g., \cite{strong2007popsyn,casanova2008,steppa2020}. More recently some authors have proposed to use the non-Poissonian spatial fluctuations in photon counts to infer the properties of unresolved sources in data \cite{bartels2016}. However, the susceptibility of this technique to systematic uncertainties in the modelling of interstellar emission seems to be sizable, therefore the results should be taken with caution \cite{leane2020}.




\section{Gamma-ray emission from the local interstellar medium: the Rosetta Stone of cosmic-ray astrophysics}\label{sec:locemiss}
Gamma-ray emission from local ($\lesssim 1$~kpc) interstellar matter is a valuable source of information to link direct CR measurements with measurements of interstellar gamma-ray emission. If the CR population was fully uniform on these spatial scales in the ISM around the Sun, then CR measurements  and gamma-ray measurements would be the expression of the same local interstellar spectrum (LIS) of CRs, and could be derived from one another based on the theories of the gamma-ray production processes and of solar modulation. In practice, given the uncertainties existing on all fronts, we can combine data and theories to get the best observational constraints and answer questions such as: how representative are direct CR measurements of the average LIS, and how much do CR densities and spectra vary within in the surrounding of the Sun and on which spatial scales?

The latter question is key to assessing the validity of some hypotheses often made in standard implementations of CR propagation models, namely that the sources are smoothly distributed in space and time, and that transport properties are homogeneous. If those hypotheses were true we would expect CR nuclei densities to vary mildly on spatial scales corresponding to the $\mathcal{O}(\mathrm{kpc})$ diffusion lengths for the GeV-TeV CR population built up over durations $> 1$~Myr. Non-uniformities in the local interstellar space therefore can inform us on the relevance of local effects on transport, and on the clustering in time and space of sources.  

\subsection{The emissivity of atomic hydrogen}

The gamma-ray emissivity per \hi atom \qhi is the most important observable quantity in this context. On one hand, \hi is diffuse and dense enough that it can probe CRs on different spatial scales with a lesser influence from localized CR sources. On the other hand, the \hi brightness temperature provides a direct estimation of gas column densities albeit the uncertainties on the optical depth correction (spin temperature).

Figure~\ref{fig:locemiss}(\textbf{a}) and~(\textbf{c}) show some recent measurements of the local \hi emissivity $q_\mathrm{LIS}$ based on \F-LAT data. \citet{casandjian2015} derived an average emissivity within a few hundred pc from the solar system by analyzing LAT data at intermediate Galactic latitudes ($10\degr < |b| < 70\degr$). We compare it to the \hi emissivity for several local clouds in three complexes that sample the same spatial region from recent studies: Chameleon \cite{plancklat2015}, the Galactic anticenter region \cite{remy2017}, and the Orion-Eridanus superbubble \cite{joubaud2020}. These three studies have been selected among the numerous results on the emissivity of local \hi \citep[including also, e.g.,][]{abdo2009,abdo2010,ackermann20113quad,ackermann2012cyg,ackermann2012ori,Ackermann:2012MC,mizuno2016,mizuno2020} for a twofold reason. 1) They employ the same sophisticated methodology to separate emission from clouds along the line of sight and from different phases of interstellar gas, which should minimize biases in the template fitting procedure due to pile-up effects and variations of \xco and dust specific opacity from cloud to cloud and within clouds themselves \cite{remy2017}. 2) They are distributed over different distances from the solar system between $\sim$150~pc and $\sim$400~pc, they lie in different directions opposite with respect to the Sun, and they span a large range of column densities and rates of star formation.

\end{paracol}
\begin{figure}[htb]
\widefigure
\includegraphics[width=1\textwidth]{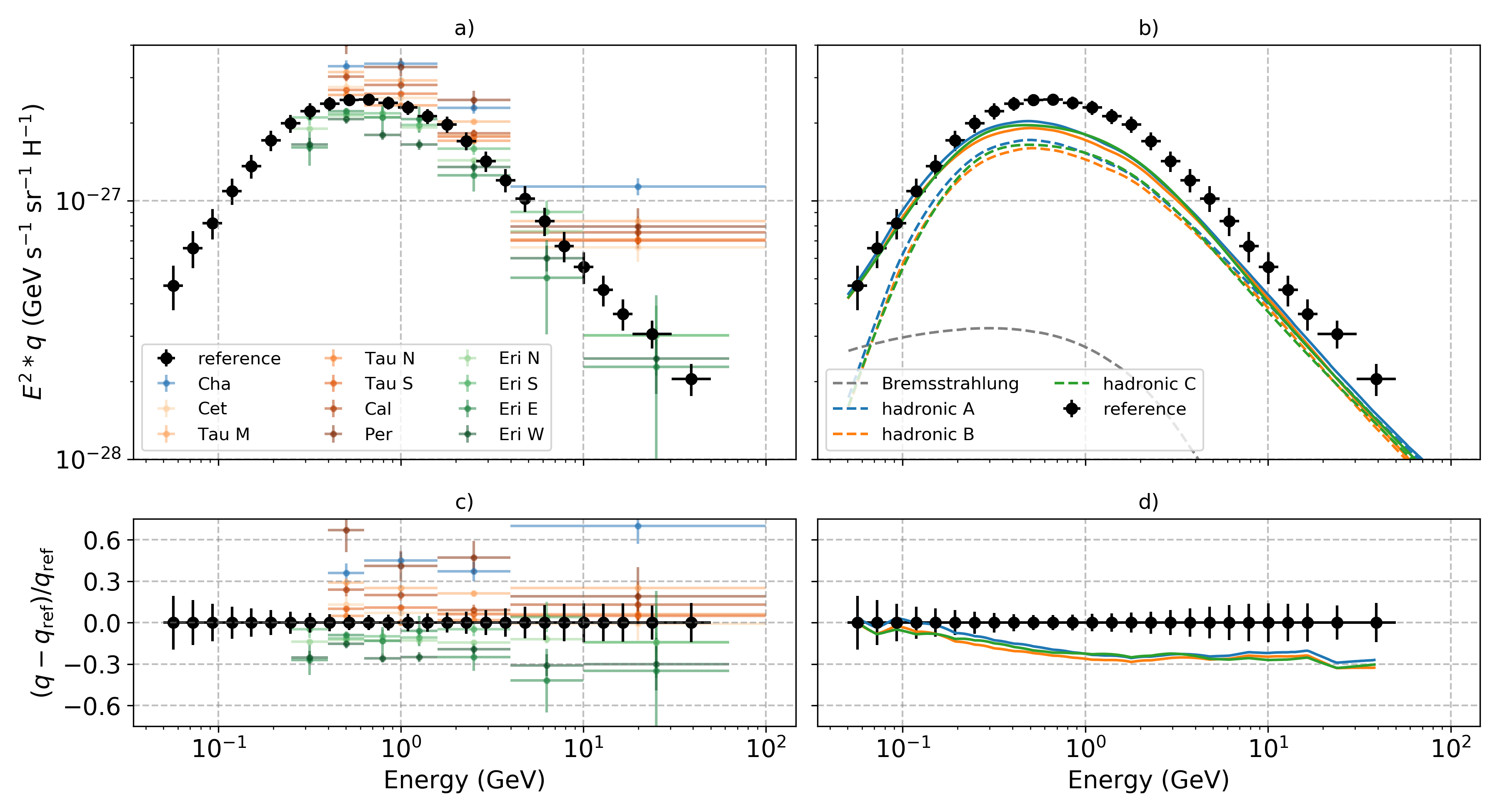}
\caption{\label{fig:locemiss} Top panels: gamma-ray emissivity per hydrogen atom in the local interstellar medium. Both panels: reference measurement $q_\mathrm{ref}$ with the \F~LAT that provides an average emissivity within a few hundred pc from the solar system \cite{casandjian2015} (combination of statistical and systematic uncertainties). (\textbf{a}) Measurements from a set of individual interstellar clouds (statistical uncertainties only) in Chameleon (blue \cite{plancklat2015}), Galactic anticenter (orange \cite{remy2017}), and the Orion-Eridanus superbubble (green \cite{joubaud2020}).
(\textbf{b}) Calculations of the emissivity based on direct CR measurements for different choices of the H and He CR spectra, and of the hadronic gamma-ray production models: model A is based on CR nuclei spectra from \cite{boschini2020} and hadronic production models from \cite{kachelriess2019} and \cite{kamae2006}; model B on CR nuclei spectra from \cite{corti2016} and the same hadronic production models as in model A; model C is based on the same CR nuclei spectra as in model A and on the hadronic production models from \cite{kamae2006} and \cite{mori2009}. In all models Bremsstrahlung is calculated based on the electron and positron LIS by \cite{orlando2018} and on the cross-section formulae by \cite{strong2000}. Hadronic emission is shown by the dashed lines, while the total including electron Bremsstrahlung is shown by the solid lines. See the text for more details on the models. Bottom panels: relative deviation of all measurements (\textbf{c}) and models (\textbf{d}) with respect to $q_\mathrm{ref}$.
}
\end{figure} 
\begin{paracol}{2}
\switchcolumn

In spite of this wide range of locations and properties, the dispersion between the different emissivity estimates is only of 25\%. The emissivity differences do not relate  to the cloud altitude above the Galactic disc, nor to the cloud location with respect to the local spiral arm. Conversely, a sizeable fraction of the difference can be attributed to uncertainties in the \hi optical depth correction. The average emissivity $q_\mathrm{ref}$ was obtained for a uniform \hi spin temperature of 140~K, while measurements from the individual regions are based on different uniform values ranging from 100~K to the optically-thin approximation (according to the best likelihood in the gamma-ray fit). When the comparison is made using the same or similar spin temperature values the differences are largely reduced \cite{plancklat2015,joubaud2020}. Another noteworthy aspect is that the emissivity of \hi is separated from other components using the template fitting technique with limitations due to modelling of other emission components and the angular resolution of the data. Therefore, at present we cannot claim any significant variations of $q_\mathrm{LIS}$ in intensity or spectrum in atomic gas within a few hundred pc around the solar system, and we conclude that the data constrain any fluctuations to $\lesssim 25\%$. In order to use gamma-ray data to probe for smaller fluctuations in the local emissivity spectrum and intensity it is crucial to reduce uncertainties in the derivation of interstellar masses, in particular in the \hi optical depth correction. Constraints could also be strengthened in the future by reducing the uncertainties in the measurement of the emissivity thanks to data with improved angular resolution and by extending it to lower energies by means of one of the proposed MeV gamma-ray missions \cite{orlando2019}.

Figure~\ref{fig:locemiss}(\textbf{b}) and~(\textbf{d}) compare the average emissivity measurement $q_\mathrm{ref}$ to calculations based on estimates of the CR LIS inferred from CR direct measurements combined with models of solar modulation and other multiwavelength constraints. We calculated the Bremsstrahlung emissivity based on the CR electron and positron LIS derived by \citet{orlando2018} using the GALPROP code tuned by using the most recent experimental results from \textit{Voyager 1} \cite{cummings2016} outside the heliosphere, from AMS-02 \cite{aguilar2014} above a few tens of GeV, as well as complementary constraints from radio, microwave, X-ray and gamma-ray emission, so that the estimate is independent of solar modulation models. As discussed by the author, the local emissivities measured by the LAT bring important constraints to the derivation of the all-electron LIS within their framework: we use the spectrum from the plain-diffusion (PDDE) model, which best fits the gamma-ray and multiwavelength/messenger data. We adopt the formulas for the Bremsstrahlung cross section given in \cite{strong2000} for electron kinetic energies $> 2$~MeV. We took into account target H and He in the ISM by assuming that the He/H fraction in the ISM is 9.6\% \cite{mori2009}.  Figure~\ref{fig:locemiss}(\textbf{b}) shows that, for the all-electron LIS considered, Bremsstrahlung is relevant only below $\sim$1~GeV and is the dominant component below $\sim$100~MeV.

In the literature there is a controversy about the compatibility of direct measurements of CR nuclei and the emissivity $\gtrsim 1$~GeV measured by the LAT, and it is not always clear wether this depends on assumptions made on CR LIS and hadronic interaction cross-sections. Therefore, for hadronic emission produced in inelastic nucleon-nucleon collisions we have considered a few different models taken from the recent literature on the subject. Model~A features the LIS of CR H, He, C, Al, and Fe derived  in \citet{boschini2020} based on data from \textit{Voyager 1} \cite{cummings2016}, PAMELA \cite{adriani2011N}, AMS-02 \cite{aguilar2015H,aguilar2015He}, HEAO-3-C2  \cite{engelmann1990} and other instruments using the GALPROP and HELMOD codes to model Galactic propagation and solar modulation, respectively. The CR LIS are used to calculate the gamma-ray emissivity per \hi atom using the H-H, H-He, He-H, He-He, C-H, Al-H, and Fe-H cross sections by \citet{kachelriess2019}, derived using the QGSJET-II-04m Monte~Carlo generator, an update of QGSJET-II \cite{ostapchenko2011} taking into account the most recent accelerator data.  The cross sections are only available above minimum energies $\geq 5$~GeV depending on the interaction considered. For H-H we combined the results from QGSJET-II-04m at proton energies $> 20$~GeV with the nondiffractive part of the cross section model by \citet{kamae2006} at lower energies owing to the result that this combination best reproduces experimental low-energy accelerator data \cite{kachelriess2012} . For other interactions, below the minimum energy available in \citet{kachelriess2019} we rescaled the H-H cross sections by the ratio of the cross section at the minimum energy available over the H-H cross section at the same energy per nucleon. We assumed again that the He/H fraction in the ISM is 9.6\%.

Model~B features the same gamma-ray production models but different H and He LIS derived by \citet{corti2016} based on a broadly overlapping set of experimental data, but a much simpler approach which consists in seeking a parametrized formula to encode the LIS and treating solar modulation by a modified rigidity-dependent force-field approximation. The resulting CR LIS differ by at most 10\% w.r.t. those by \cite{boschini2020} in the rigidity range 1-10~GV where solar modulation still plays a significant role but there are no direct constraints on the LIS from \textit{Voyager}. This translates to a $\lesssim 5\%$ difference in the resulting gamma-ray emissivity spectrum below a few GeV after convolution with the gamma-ray production cross-sections. An additional source of difference between models A and B is that in model B we neglect the interactions from CR species heavier than He, which contribute $\sim$1\%-3\% of the emissivity in model A.

Model~C employs the same CR H and He LIS spectra as in model~A, but different gamma-ray production models, namely the H-H cross sections by \citet{kamae2006}, including the diffractive part, and the scaling factors by \citet{mori2009} based on the DPMJET-3 Monte~Carlo generator to scale the H-H emissivity in order to account for heavier elements, namely CR He nuclei and He, CNO, Mg-Si, and Fe in the ISM. The scaling factors are available only for CR energies of 10~GeV/nucleon, therefore we neglect their variations with energy, that should be modest at least in the energy range $>10$~GeV \cite{kachelriess2014} but are expected to become more relevant in the less understood energy range $<10$~GeV (see Figure~1 of \cite{mori2009}, which only shows the results down to energies of 5~GeV/nucleon, below which results become inaccurate). We note that the contribution from target nuclei heavier than He, not accounted for in models~A and~B, represents $\sim$2\% of the total emissivities in model~C. 

To summarize, in the hadronic-dominated energy range $>100$~MeV we find variations between different predictions of the emissivity based on direct CR measurements of $\lesssim 10$\%, more important at lower energies. The differences are comparable to the uncertainties in the most precise emissivity measurements and a factor of $\sim$2 smaller than the dispersion between measurements for different individual clouds.

For gamma-ray emission in the energy range $>10$~GeV, where the differences between alternative estimations of the CR LIS and between alternative hadronic production models are small, the average emissivity $q_\mathrm{ref}$ exceeds the predictions by 20-30\%. This broadly agrees with several independent results in the literature \cite{dermer2013,strong2015,orlando2018}. The recent precise measurements by AMS-02, in particular of He \cite{aguilar2015He}, rule out earlier claims  \cite{casandjian2015}  that the average emissivity measured by the LAT is consistent within uncertainties with direct CR measurements as pointed out by \citet{orlando2018}. We also stress that the evaluation of the contributions from species heavier than H must take into account these measurements, which, for instance, is not necessarily the case when using {\it enhancement factors} taken from the literature such as the one provided by \cite{mori2009}. The latter aspect is discussed in greater detail by \cite{kachelriess2014}.

However, the difference between the predictions and the average emissivity is comparable to the dispersion between different clouds, that, as discussed above, is largely imputable to uncertainties in the \hi optical depth correction performed assuming a uniform spin temperature and possibly other analysis features. Once the dispersion of $\sim$25\% among different regions and clouds is taken into account, regardless of its interpretation as a systematic uncertainty related to the extraction of the emissivity or as an effect of real fluctuations in CR densities on spatial scales smaller than a few hundred pc, for the moment it should still be considered as viable that direct CR measurements are in agreement with the local \hi gamma-ray emissivities.

Therefore, also to connect direct CR measurements and gamma-ray observations of the local ISM it is crucial to reduce uncertainties in the extraction of the emissivities and derivation of interstellar masses. At the same time, to reduce uncertainties in the derivation of the gamma-ray emissivities from direct CR measurements it is paramount to improve our knowledge of the gamma-ray production cross sections, especially at low energies reaching the kinematic threshold for pion production where currently only approximate scaling of H-H is readily accessible to account for interactions of heavier species. To this end it would be extremely useful to pursue experimental accelerator campaigns going beyond those already undertaken, to further improve Monte~Carlo generators, and to make more results \citep[e.g.,][]{mazziotta2016} publicly available in the form of parametric/tabulated cross sections for gamma-ray production. 

\subsection{Molecular clouds and their spectra}\label{sec:clouds}

An alternative approach consists in studying gamma-ray emission from molecular clouds. Molecular clouds provide localized (few tens of pc) targets for CR interactions with large masses. However, with observations of molecular clouds two additional classes of complications arise.
\begin{enumerate}
\item The column densities of molecular gas are traced only indirectly, most often via the CO lines or dust thermal emission. On one hand, the \xco ratio is empirically observed to vary by a factor of $\sim$2 between different local clouds depending on their diffuseness \cite{remy2017}. Furthermore, based on numerical simulations \xco is expected to vary significantly within individual clouds between the inner part that is more self-shielded and the outskirts where CO is more easily photodissociated by UV photons \cite{bertram2016}. On the other hand, based on gamma-ray and multiwavelength data the conversion factor from dust thermal emission to gas column density is now well-established to vary by a factor of $\sim$3 from the low-column densities in atomic gas to cold and dense gas in molecular cloud cores probably owing to evolution of the dust grain properties (chemical composition due to irradiation and structure due to irradiation itself, but also mantle accretion, coagulation, and ice coating) \cite{remy2017,remy2018}. Therefore, uncertainties of a factor of a few affect the estimates of the absolute level of CR densities from gamma-ray emission of molecular gas, which appears better suited to probe the spectral shape of CRs.
\item Two different kinds of physical phenomena can affect the spectrum of the CR populations around molecular clouds so that it is no more representative of the large-scale population. On one hand, molecular gas is associated with star formation, which in turns triggers all the phenomena expected to accelerate particles efficiently. Therefore, freshly-accelerated particles can be observed in this environment (see Section~\ref{sec:youngcr}). In the literature molecular clouds free from the influence of nearby accelerators are often referred to as \emph{passive clouds}, but establishing which clouds belong to this category observationally is challenging. On the other hand, several effects may alter the propagation of CRs in molecular clouds in an energy-dependent fashion. Increased ionisation losses were the first mechanism considered historically that could suppress CR densities below 300~MeV kinetic energies \cite{skilling1976}, but nowadays they are not considered by themselves the most important effect. Streaming instabilities driven by CR pressure gradients can alter the CR transport regime from diffusive to advective and so suppress the CR pressure by $\lesssim 10\%$ at GeV energies \cite{everett2011}. The peculiar magnetic field configuration in molecular clouds via magnetic mirroring coupled with energy losses can lower the CR densities by a factor 2-3 in molecular cores in the MeV energy range, while CR enhancements due to magnetic focussing appear a less important effect \cite{padovani2011}. Damping of magnetic turbulence by ion-neutral friction coupled again with energy losses can reduce the CR fluxes below 100 MeV kinetic energies, especially for electrons \cite{phan2018}.  We note that all propagation effects mainly affect CRs at energies $< 1$~GeV, for which constraints from gamma-ray observations are looser.
\end{enumerate}

To this date there is a controversy in the literature about the uniformity of the spectral shape of gamma-ray emission from molecular clouds in the local interstellar space as observed with the \F~LAT. Although most studies agree on the uniformity of the spectra and their compatibility with the spectrum of \hi, \citet{yang2014} claimed low-energy CR enhancements in the Orion A, Orion B, and Chameleon clouds, not confirmed by subsequent studies \cite{plancklat2015,neronov2017}. More recently, \citet{baghmanyan2020} claimed spectral deviations with respect to the LIS for the molecular clouds in the Aquila rift, rho Ophiuchi, and Cepheus. For the latter two the results are at odds with previous studies based on smaller datasets \cite{abdo2010,Ackermann:2012MC,neronov2017}. The devil seems to lie in the detail, that is, how the different studies account (or not) at the analysis level for variations of CR densities in different structures along the line of sight and in the Galactic background, and variations/uncertainties in \xco and/or dust-to-gas conversion factors. Analysis assumptions about all of these aspects combined with the energy-dependent PSF of the LAT may easily produce distortions in the derived spectra via cascade effects in the template fitting procedure. The claims of spectral variations therefore require further investigation to asses their robustness.

By stacking the spectra of several nearby molecular clouds measured by the \F-LAT \citet{neronov2017} could highlight the existence of a break in the CR nuclei spectrum at a rigidity of $18^{+7}_{-4}$~GV, which is consistent with the direct measurements from \textit{Voyager 1} \cite{cummings2016} and AMS-02 \cite{aguilar2015H,aguilar2015He} in a rigidity range where the impact of solar modulation is most relevant. \citet{neronov2017} speculated that this coincides with the transition from the Galactic-scale steady-state population observed throughout the disk of the Milky Way (see Section~\ref{sec:largescale}) to a local population of CRs driven by stochastic injection around the Sun localized in space and time. We note, however, that uncertainties in the hadronic cross-sections have not been taken into account in this work.

First results at TeV energies were recently published by the HAWC collaboration \cite{hawc_mc}. No detections are reported for a set of seven nearby molecular clouds. Upper limits on their gamma-ray fluxes are less than an order of magnitude larger than expectations based on the extrapolation of direct measurements by AMS-02. For a stacked analysis assuming a power-law CR spectrum with index~2.7 the upper limit on the CR density is approximately at the level predicted by direct measurements. This implies that, if the CR spectrum follows a simple extrapolation of what is directly measured at lower energies, HAWC should reach a detection of nearby clouds by doubling the exposure, or even faster taking into account the ongoing detector upgrades. Complementary results can be expected from LHAASO and, eventually, from SWGO (although most of the nearby molecular clouds in the Gould Belt are visible from the Northern hemisphere).

\section{Large-scale interstellar gamma-ray emission: tracing cosmic rays throughout galaxies}\label{sec:largescale}

The large-scale distribution of CRs in galaxies encodes information on their injection sites and spectra, and on the transport mechanisms and their interactions with other components of the ISM. For a long time it has been known that gamma-ray data point to the existence of a large-scale population of CRs in the Milky Way disk with properties similar to that observed near the Earth, see, e.g., \cite{strong1988}, and of CR populations with somewhat different properties in the Magellanic Clouds \cite{sreekumar1992,sreekumar1993}. Recent years have seen impressive developments in GeV observations for the Milky Way and local-group galaxies. 
Their CR populations show diverse and sometimes unexpected properties. Variations in the densities and spectra of CRs across the galactic disks and halos with departures from expectations based on standard implementations of CR models, combined with ubiquitous unexplained residual  features, are questioning our understanding of particle transport and its microphysical foundation. At the same time, the first sub-MeV and TeV observations have been opening new and complementary windows to constrain the large-scale CR distribution for the Milky Way.


\subsection{Cosmic-ray distributions through the disks of galaxies}\label{sec:disks}

\subsubsection{The Milky Way}\label{sec:diskmw}

A lot of recent progress is based on data from the \F~LAT. A first avenue to infer the large-scale CR distribution from the observations is through the emissivity of interstellar gas, that can be extracted using the template-fitting technique for multiple regions along the line of sight thanks to the distance proxy provided by the Doppler shift of atomic or molecular lines (Section~\ref{sec:modda}). This approach has been applied to two regions towards the outer Galaxy in the second and third Galactic quadrants for which the separation of different structures and spiral arms along the line of sight is remarkably good and free from ambiguity \cite{abdo2010,ackermann20113quad}. These studies strengthened the observational evidence for the long-known \emph{gradient problem} \cite{strong1988}, i.e., the fact that the gamma-ray emissivity only mildly decreases from the position of the Sun up to Galactocentric radii of $\sim$15 kpc, while the number density of putative CR sources shows a steep decline. Furthermore, in the third quadrant the data allow only a decrease $<20\%$ in the low-density region between the local and Perseus arms, ruling out a strong coupling between CR and ISM densities invoked by some early modelling efforts \cite{bignami1975}.

The distribution of CRs throughout the entire Galactic disk was derived by several authors by analyzing LAT data for the entire sky with gas templates built by separating \hi and CO in Galactocentric rings \cite{Acero:2016qlg,Yang:2016jda,Pothast:2018bvh} and using the \hi emissivity to infer the underlying CR spectra. These studies show that the CR proton density above 10 GeV varies by a factor of a few across the disk of the Milky Way, with a peak at $\sim$4~kpc from the Galactic center and a mild decrease as a function of radius beyond 5 kpc, which confirms and extends the trend inferred from the dedicated studies of the outer Galaxy. More surprisingly, LAT data have revealed a progressive hardening of the CR proton spectrum toward the inner Galaxy, with the power-law spectral index at 10~GeV going from the local value of $\sim$2.7 near the Sun to $\sim$2.5 at a few kpc from the Galactic center. For these studies, as well as for the outer Galaxy, a major source of uncertainty is due to the \hi opacity correction (spin temperature), which is associated with a $\sim$30\% systematic uncertainty in the \hi emissivities \cite{ackermann20113quad,Acero:2016qlg}. Another important limitation is the possible contamination  by diffuse emission from populations of unresolved sources. The fraction of unresolved sources in the total diffuse emission is estimated to be 3\% at 1 GeV \cite{3fgl}. We expect this fraction to increase as a function of energy, but based on reasonable assumptions about the source populations the conclusions on the general trends in CR densities and spectra inferred from LAT data remain unchanged  \cite{Pothast:2018bvh}. Other sources of uncertainty include the modelling of IC emission and detected gamma-ray sources.

Furthermore, the impact of the assumption that CR densities are axisymmetric, and, in particular, that the near and far region within each ring for radii smaller than the solar circle share the same CR densities, needs to be checked against observations not affected by the kinematic ambiguity. A complementary approach that may overcome this limitation is the use of well-localized targets. \citet{aharonian2020mc} have derived the emissivity of nineteen giant molecular clouds located at Galactocentric distances up to 12~kpc. The trend in inferred CR proton densities as a function of Galactocentric radius is generally consistent with earlier works based on ring emissivities \cite{Acero:2016qlg,Yang:2016jda}, but the large uncertainties due to the separation of the clouds from foreground/background gas and the conversion of CO intensity to \hd column density makes the results too uncertain to draw robust conclusions. A few clouds show spectral deviations from the general trend that may point to localized CR excesses, e.g., due to a nearby accelerator. More recently the same authors analysed LAT observations of nine clouds located at Galactocentic distances of 1.5-4.5~kpc employing dust opacity (inferred from thermal emission) as total gas tracer (no separation along the line of sight), and they obtained results at odds with the ring analyses \cite{peron2021}. However, we warn the readers that the latter results are based on the assumption of a constant dust specific opacity across the Milky Way with an uncertainty of 20\%, while local clouds show that variations of a factor $\sim$3 related to evolution of the dust grain properties are possible  (Section~\ref{sec:clouds}), and, moreover, an increase of the dust specific opacity of a factor of a few in the inner Galaxy is expected from the correlation of dust-to-gas ratio and metallicity gradient as a function of galactocentric radius observed for external galaxies \cite{issa1990,boissier2004,munoz2009}.  Different measurements of the CR proton density and spectrum across the Galactic disk are summarized in Figure~\ref{fig:crgrad}.

A second avenue to constrain the large-scale CR distribution consists in comparing the data directly to the outcome of predictive models. \citet{Ackermann:2012galprop} compared predictions by GALPROP to the LAT data for the entire sky by varying hypothesis on the astrophysical input such as CR source distribution or \xco. The results are consistent with the trends seen in the template analyses, notably the flat CR profile in outer Galaxy, and higher/harder emission toward the inner Galaxy. Although the data/model agreement is reasonable and demonstrate that standard implementations of Galactic CR propagation models describe the gamma ray data within $\sim$30\%, regions of extended residuals appear for any of the models considered.  Residual features will be discussed later in Section~\ref{sec:residuals}. The study \cite{Ackermann:2012galprop} also demonstrates the high level of degeneracy between different inputs to predictive models, and therefore the importance to use different complementary approaches to analyse and interpret the data.

Predictions from GALPROP were also compared to data from \textit{INTEGRAL}~SPI, which reveals the existence of diffuse emission from the inner $60\degr$ of the Galactic disk in the soft gamma-ray band from 20~keV to 2 MeV. The diffuse gamma-ray emission above 60~keV is consistent with a dominant origin from inverse-Compton scattering of CR electrons , and connects spectrally with emission measured at higher energies by COMPTEL and pair-conversion telescopes \cite{porter2008,bouchet2011,churazov2020}. These data tend to favor models with an important contribution from secondary electrons and positrons, which is not necessarily the case for the electron spectrum measured near the Earth \cite{orlando2018}. A new space mission for MeV astronomy holds the potential to deepen our understanding of the large-scale properties of IC emission and bridge \textit{INTEGRAL} and \F observations at a sensitivity largely improved compared to COMPTEL \cite{orlando2019}.

Recently we gathered the first measurements of diffuse emission in the energy range from hundreds of GeV to 1~PeV in different regions of the Galactic disk thanks to Milagro \cite{abdo2008}, H.E.S.S. \cite{abramowski2014}, ARGO-YBJ \cite{bartoli2015}, HAWC \cite{zhou2017}, and Tibet AS$\gamma$ \cite{amenomori2021}. In this energy range the key challenge, beside separating diffuse gamma-ray emission from the charged-particle background, is to disentangle the interstellar component from the contributions from unresolved source populations. For example, \citet{steppa2020} estimate that $\sim$30\% of the diffuse emission measured by H.E.S.S. can be attributed to unresolved source populations, and \citet{amenomori2021} estimate the same fraction to be 13\% for the measurement at energies above 100~TeV with Tibet AS$\gamma$. While the measurements generally exceed expectations based on the local CR spectrum and there is now firm evidence of emission from CRs up to the knee, there are tensions between observations with different instruments \cite{Neronov:2020} and better constraints on the unresolved source contribution become key to use TeV data to investigate CR properties such as the spectral hardening in the inner Galaxy \cite{pagliaroli2020}. At the same time, upcoming measurements with CTA and LHAASO \cite{neronov2020iact,neronov2020lhaaso}, and, possibly, SWGO will provide much improved sensitivity, and enable an even higher complementarity with neutrino measurements. 


\subsubsection{Implications of the gradient problem and inner-Galaxy hardening}\label{sec:gradhard}

\end{paracol}
\begin{figure}[htb]
\widefigure
\includegraphics[width=0.95\textwidth]{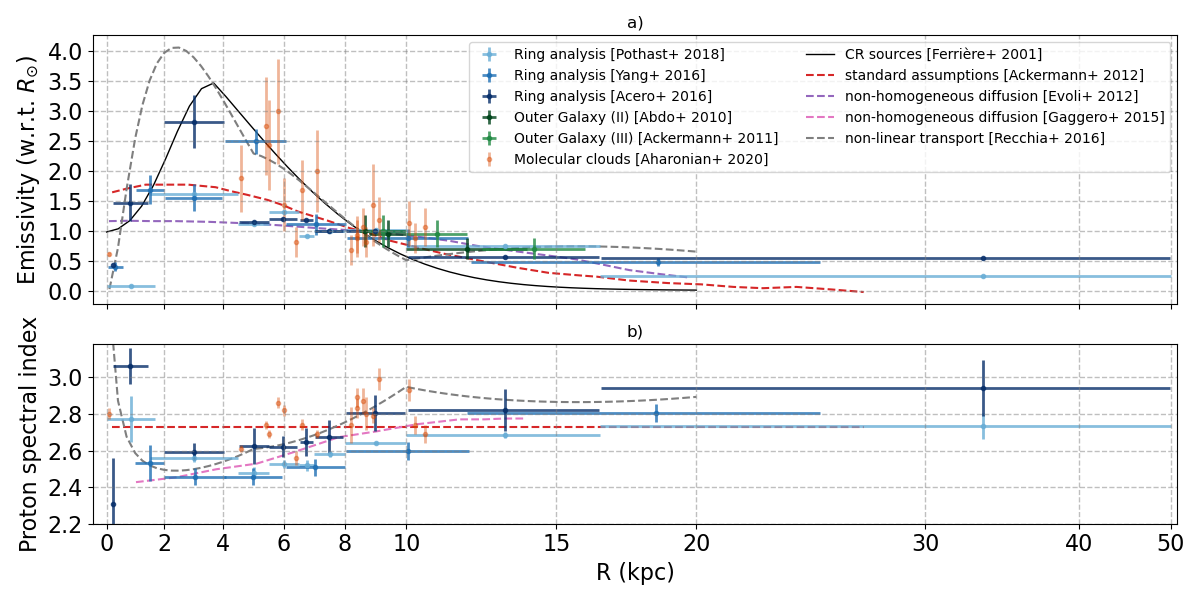}
\caption{\label{fig:crgrad} Top panel: gamma-ray emissivity/CR proton density as a function of Galactocentric radius derived for the entire Milky Way using ring analyses \cite{Acero:2016qlg,Yang:2016jda,Pothast:2018bvh}, for the outer Galaxy \cite{ackermann20113quad,abdo2010}, and for a sample of giant molecular clouds \cite{aharonian2020mc}. Values are normalized for each set of measurements by the value for the region including the solar circle ($R_\odot =8.5$~kpc). For giant molecular clouds the normalization is the average for all clouds in the Gould Belt ($R$ between 8.2~kpc and 9.1~kpc). Bottom panel:  CR proton spectral index as a function of Galactocentric radius derived by a subsample of the analyses shown in the top panel. We remark that the gamma-ray energy range considered varies between the different analyses, but whenever readily available we show in the plot the inferred CR proton density and spectral index at 10 GeV. In both panels we include predictions from some of the models discussed in the text, namely a standard CR model implementation \cite{Ackermann:2012galprop} (red), two models with non-homogeneous diffusion tuned to reproduce the CR density profile in the outer Galaxy \cite{Evoli:2012ha} (purple) and the inner-Galaxy hardening \cite{Gaggero:2014xla} (pink), and the non-linear transport model by \citet{Recchia:2016bnd} with exponential cutoff of the magnetic field strength at large $R$ (grey). In the top panel we also show the putative CR source profile taken from \cite{Ferriere2001} (black) as useful reference.}
\end{figure} 
\begin{paracol}{2}
\switchcolumn

The gradient problem and inner-Galaxy hardening observed by \F have been stimulating a lively debate on our understanding of CR transport in the Milky Way. As far as the gradient problem is concerned, the assumption of a very extended $\sim$10~kpc diffusive halo seems to alleviate the discrepancy. However, this solution is in tension with the most recent observations of CR isotopic abundances and, possibly, radio/gamma-ray emission from the Milky Way, as discussed in Section~\ref{sec:halo}, and also covered for instance in \citeauthor{Evoli:2012ha} \cite{Evoli:2012ha} and references therein. Also, ad-hoc assumptions on the distribution of CR sources may contribute to mitigate the problem, and suffer from similar problems when confronted with catalogs of SNR and pulsar, that are expected to trace the radial distribution of the CR injected power.

A promising attempt to solve the gradient problem based on non-homogeneous diffusion, also in connection with the anisotropy problem (i.e. the larger dipole anisotropy predicted by standard implementations of the Galactic CR transport model with respect to the observed one) was presented in \citet{Evoli:2012ha}. In that paper, a correlation between the CR source density and the normalization of the diffusion coefficient is invoked. The proposed solution (visualized in Fig. \ref{fig:crgrad}, upper panel, ``non-homogeneous diffusion'') stems from the following physical argument: a larger turbulence level is expected in the regions of the Galaxy characterized by a larger density of CR accelerators, in particular, along the Galactic plane, in the range of Galactocentric radii close to the so-called molecular ring.  Given the topology of the large-scale regular magnetic field, and the overall geometry of the problem, the CRs accelerated in the Galactic plane mainly escape in the vertical direction, perpendicular with respect to the regular field. Given the increase of the perpendicular diffusion coefficient with increasing turbulence level that is highlighted in several numerical simulations (see for instance \cite{DeMarco:2007eh,Snodin:2015fza,Dundovic:2020sim}), the aforementioned correlation naturally follows from these considerations.

As far as the hardening problem is concerned, a phenomenological model where the trend is obtained as a result of a smoothly varying scaling of the diffusion coefficient with respect to rigidity was presented in \citet{Gaggero:2014xla} (see Fig. \ref{fig:crgrad}, lower panel, ``non-homogeneous diffusion'').
A physical interpretation of this trend in terms of specific aspects of transport physics was recently presented in \citet{Cerri:2017joy}. In this analysis, the argument is once again based on the nature of perpendicular escape and the geometry of the magnetic field. 
The starting points are the following considerations: 
{\it (i)} the numerical simulations that aim at characterizing CR transport in pre-exisisting (Alfv\'enic) turbulence, already mentioned above, suggest a different scaling with rigidity as far as parallel and perpendicular diffusion coefficients are concerned, with the parallel transport featuring a harder rigidity dependence;
{\it (ii)} the state-of-the-art models of the large-scale Galactic magnetic field (see for instance \cite{janssonfarrar2012}) seem to suggest the presence of an X-shaped poloidal component in the inner Galaxy; hence, the vertical escape of CRs may be parallel in the inner Galaxy, and perpendicular in the outer Galaxy, where the field is expected to follow the spiral pattern on the Galactic plane, with a less prominent vertical component.
These facts imply a progressively harder scaling of the propagated CR spectral index, as simulated by \cite{Cerri:2017joy} with an axisymmetric, fully-anisotropic version of the DRAGON code.

Following a very different line of thought, another recent work \cite{Recchia:2016bnd} attempts to explain both the gradient and the spectral hardening problems at the same time. In this work the non-linear effects mentioned in Section~\ref{sec:modda} are exploited to provide an explanation to both anomalies. The idea is that the regions with a larger density of accelerators feature a larger CR gradient. Hence, the turbulence growth rate associated to streaming instability is larger: as a consequence, the non-linear phenomenon of CR self-confinement is greatly enhanced. In the GeV domain CR escape is actually the result of the competition between rigidity-dependent (possibly self-generated) diffusion and rigidity-independent advection. The more efficient self confinement implies a lower diffusion coefficient associated to self-generated turbulence. Hence, advection takes over up to larger rigidities, and the propagated spectrum is less affected due to the energy-independent nature of this process. Hence, the inner regions of the Galaxy are expected to feature a harder CR spectral slope closer to the one initially injected in the ISM. A careful numerical treatment of this problem actually shows that the gradient problem may also be fixed, provided that the magnetic field is assumed to drop exponentially at large radii. The model is represented in Fig. \ref{fig:crgrad}, both lower and upper panel (``non-linear transport'').  

The interpretation of the spectral hardening in terms of the interplay between advection and non-linear CR confinement has an important consequence. This kind of solution is valid only in the low-energy range where the streaming instability plays a major role, and cannot be invoked if the spectral trend were to be clearly confirmed at rigidities $\gtrsim 100$ GV. The current data do not allow to reach a firm conclusion on this point. However, the analysis presented in \cite{Pothast:2018bvh}, already discussed above, seems to point towards the presence of a spectral hardening even in high-energy \F-LAT gamma-ray data. 
On the other hand, the explanations based on the different scaling relations of perpendicular and parallel transport are expected to hold at all energies.

The different explanations of the inner-Galaxy hardening are expected to result in different phenomenological implication for the multi-TeV gamma-ray interstellar emission. In particular, a harder hadronic CR spectrum towards the Galactic Center that extends up to the multi-TeV domain (not covered by \F-LAT data) would imply a significantly larger diffuse gamma-ray emission from the inner Galactic plane. 
Different scenarios were put forward in order to bracket the uncertainties in this context, mainly due to the extrapolation of the analyses based on Fermi-LAT data, and the most ``optimistic'' prediction were shown to saturate the diffuse emission from the Galactic Ridge measured by H.E.S.S., leaving less room for the contribution from a central accelerator  \cite{Gaggero:2017jts}. 
These scenarios featuring the hardening in the multi-TeV domain are compatible with the interpretation based on the idea of anisotropic transport, and show some tension with the interpretation that stems from the non-linear propagation models, given that self-generated diffusion models predict a transition to transport dominated by background turbulence at $\sim$100~GeV, while, as mentioned before, the peculiar scaling relations associated to anisotropic transport extend to larger energies.
However, more data and further studies on the modelling sides are needed. Currently operating and forthcoming air shower and Cherenkov experiments (in particular LHAASO, HAWC, SWGO, and CTA)  will help to shed light on this issue. In particular, if the presence of a harder spectrum in the inner Galaxy were to be confirmed in the TeV-PeV range, the  interpretation of this effect in terms of the interplay between advection and self-confinement would be disfavored.
On the modelling side, as pointed out for instance in \cite{Lipari:2018gzn}, more advanced simulations of the multi-TeV gamma-ray emission from the Galactic plane will be needed, also taking into account the crucial effect of absorption that plays a dominant role especially above 50-100 TeV.

We remark that, if confirmed, the presence of a progressive spectral hardening of the hadronic CR population in the inner Galaxy has also interesting phenomenological consequences in a multi-messenger context. In fact, as pointed out for instance in \cite{Gaggero:2015xza}, phenomenological models that reproduce this trend by featuring a radially-dependent slope of the diffusion coefficient predict a neutrino flux in the inner Galaxy that is 2-5 times larger compared to conventional predictions, based on a constant spectrum across the Galactic plane (see also \cite{Pagliaroli:2017fse}). This may explain up to 25\% of the neutrino events measured by IceCube, and may suggest that a detection of a positive correlation of the IceCube events with the Galactic plane might be just around the corner. This hypothesis was extensively tested by the ANTARES and IceCube Collaborations. The recent analysis  \cite{Albert:2018vxw} provided joint constraints that start to challenge this scenario. The constraints on a Galactic component are based on 10 years of ANTARES showers and tracks (218 showers in total), and 7 years of IceCube tracks (730130 events with 191 events expected from the optimistic Galactic model). The results are in mild tension with the most ``optimistic'' versions of these phenomenological models, and allow for the possibility to test this kind of prediction in the near future. A more recent analysis \cite{Aartsen:2019epb} based on seven years of IceCube cascade data (characterized by an interaction vertex inside the detector) outlined a $2\sigma$ hint for a Galactic component consistent with the optimistic models. Future studies are needed to shed light on this issue. A firm detection of a Galactic neutrino component would represent a remarkable confirmation of the presence of a hard hadronic CR population extending above the TeV domain, and would greatly help in shedding light on the microphysics processes that originate this anomaly.

\subsubsection{Local-group galaxies}\label{sec:localgroup}

Moving to external galaxies, the LMC is the best target when it comes to studying how gamma-ray emission connects to components of a galactic ecosystem due to its proximity (distance of $\sim$50\,kpc) and a favourable geometry (disk-like structure with a low inclination angle $\sim30^{\circ}$).  The emission at 1\,GeV is dominated by radiation seemingly correlated to the gas disk, as could have been expected based on what we see in the Milky Way. The gamma-ray distribution in the disk could be fitted assuming an emissivity profile decreasing by a factor 2-3 from the center to the outskirts, with a peak value of about 30\% of the emissivity measured in the Solar system neighbourhood \citep{Ackermann:2016}. This lower value is thought to arise from the smaller size of the LMC, and a corresponding smaller confinement volume. Again, this is very reminiscent of the Milky Way (see above in this section). More surprising, though, is the fact that the emission at 10\,GeV is contributed at about 50\% by extended components of unknown origin but not evidently correlated with gas or recent star formation, therefore implying localized enhancements in the CR densities by factors of 2 to 6, or an alternative explanation not related to interstellar emission \citep{Ackermann:2016}. To date, the origin of such features is still unexplained, and revisiting the gamma-ray emission from the LMC from the twice larger Fermi-LAT data set available now and increased exposure with H.E.S.S. would be timely.

The SMC is another promising target due to its proximity (distance of $\sim$62\,kpc) but the geometry of the galaxy is much more intricate, with an irregular shape elongated along the line of sight over 20\,kpc \citep{Scowcroft:2016} that complicates the interpretation of observations. Significant gamma-ray emission is detected but no obvious correlation with gas or star formation is observed \citep{Abdo:2010smc,Caputo:2016}. If interpreted anyway as interstellar emission, the flux observed implies an average CR density of about 15\% of the emissivity measured in the solar system neighbourhood. However, it was estimated that the measured flux could be accounted for to a large fraction by an unresolved population of pulsars, which would imply an even lower CR density \citep{Abdo:2010}.

At a much larger distance of 785\,kpc, another target of choice is M31 (Andromeda) that, as a grand design spiral galaxy, resembles more closely our Milky Way and may allow a more direct analogy. Extended emission from the galaxy is detected but the signal is confined to the inner regions, within 5\,kpc from the center. It does not fill the disk of the galaxy and in particular does not correlate with the regions rich in gas or star formation activity that are mainly located in a large ring at 10\,kpc distances from the center. Emission from the disk at large is however not strongly excluded and may be present at a level of up to 50\% of the currently detected flux \citep{Ackermann:2017m3133}. The different gamma-ray emission distribution in M31, compared to the Milky Way, can be interpreted as resulting from its global properties: the star formation rate in M31 is about 10 times lower than in the Galaxy, which would decrease the contribution from the star-forming disk, while it has a 5-6 times more massive bulge, which would enhance the contribution from old stellar populations gathered in the central regions.

\subsection{Cosmic-ray distributions through the halos of galaxies}\label{sec:halo}
As briefly described in Section \ref{sec:crtheory}, within the standard galactic CR paradigm it is assumed that particles are confined diffusively in a magnetized halo with a size of a few kpc for the Milky Way. 
Yet, the formation of the halo and the escape at its boundary are still poorly understood, and usually modelled by imposing free escape at a fixed height $z_{max}$. The value of $z_{max}$ is then treated as free parameter in CR models adjusted to reproduce CR elemental/isotopic abundances as well as the spectra of CR species measured locally. Recent measurements with AMS-02 and other instruments in this framework point to values of $z_{max}=5$~kpc with sizeable uncertainties of a few kpc \cite{Evoli2020,Weinrich2020}.

Until recently, gamma-ray and radio emission from CR interactions was used to constrain the propagation in the Milky Way halo only indirectly via aggregate properties (longitude, latitude, and radial profiles and spectra), which is tricky due to severe degeneracies with other unknowns such as the source distribution, e.g., \cite{Ackermann:2012galprop, orlando2013}. Conversely, kpc-wide synchrotron halos around external edge-on galaxies have been observed in radio for almost thirty years, e.g., \cite{dahlem1994}.

The last decade has seen two major advances in the understanding of the halo and related observational constraints especially from gamma rays: more direct observations in gamma rays thanks to observations with the LAT of clouds at large distances from the Galactic plane, and, possibly, of the halo of M31; the emergency of models of galactic CR halos based on the microphysics of magnetic turbulence and particle transport.

The improved sensitivity of the LAT made it possible to measure the emissivity from atomic clouds at large distances from the disk. Those include notably a set of high- and intermediate-velocity clouds which span heights from a few hundred pc to several kpc \cite{tibaldo2015} with robust distance determinations based on stellar brackets \cite{wakker2001}. Their emissivities, shown in Figure~\ref{fig:vertical_gradient}, testify to a significant decrease of CR densities as a function of distance from the Galactic plane. On one hand, this represents a new and robust observational test of the Galactic origin of CRs below the knee. On the other hand, the emissivity measurements can be compared to predictions from CR propagation models and directly constrain the CR gradient in the halo. A fit to these measurements yields a $z_{max}$ value of $\sim$2~kpc with an uncertainty of a few kpc \cite{tibaldo2015}. In particular, the upper intermediate-velocity Arch at a height of 0.7-1.7~kpc above the disk was found to have an emissivity $<45\%$ relative to the local value at 95\% confidence level.

\begin{figure}[htb]
\centering
\includegraphics[width=0.5\textwidth]{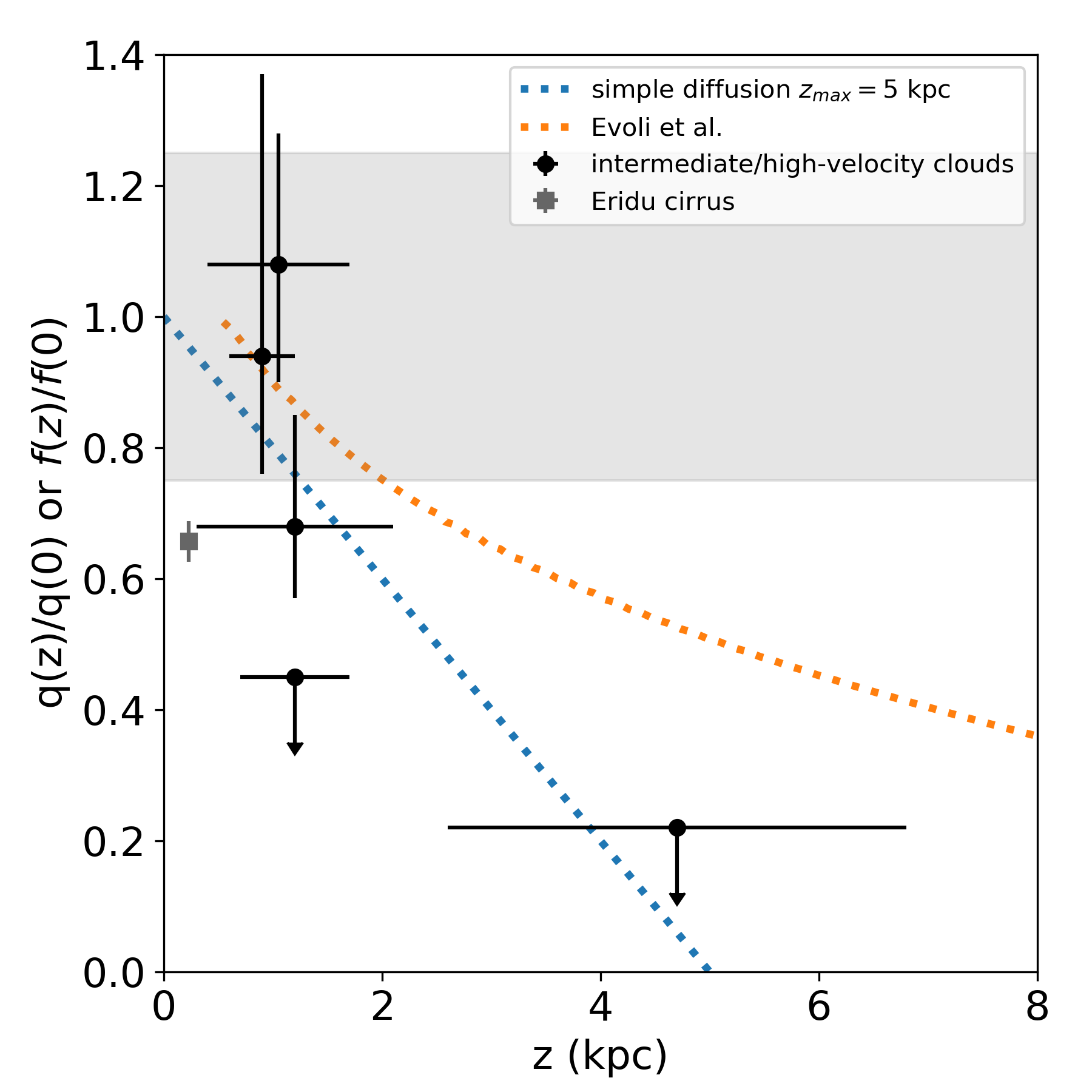}
\caption{\label{fig:vertical_gradient} Vertical gradient of gamma-ray emissivities or cosmic-ray densities in the Milky Way halo. The points correspond to emissivity measurements derived from \F~LAT data for intermediate- and high-velocity clouds with distance brackets based on stellar probes \cite{tibaldo2015} and for the Eridu cirrus \cite{joubaud2020}. The horizontal band shows the 25\% dispersion of emissivity measurements for nearby clouds (Section~\ref{sec:locemiss}). The lines correspond to CR densities from two models: the z-dependent part of the solution of the diffusion equation in the plane-parallel geometry (infinitely thin Galactic plane) with a uniform source distribution when only ionization losses are assumed for the halo height $z_{max}=5$~kpc inferred from recent direct CR measurements \cite{Evoli2020,Weinrich2020} and the model by \citet{Evoli:2018nmb} which describes CR vertical propagation based on a mixture of advected turbulence and of CR self-generated waves (we show the CR densities at 10~GeV which are most representative for the gamma-ray energies considered).}
\end{figure}   

Even more recently, \citet{joubaud2020} reported an emissivity of $0.657 \pm 0.031$ with respect to the local value for the Eridu cirrus at a modest altitude of 200-250~pc (also shown in Figure~\ref{fig:vertical_gradient}). We note that the distance to the Eridu cirrus is based on dust reddening \cite{joubaud2019}, a method more indirect than the use of stellar brackets. For the moment it remains unclear whether this observation should be interpreted in terms of the large-scale vertical gradient of CRs or a peculiar magnetic field configuration in this cloud \cite{joubaud2020}.

It is therefore essential to study in gamma rays a broader sample of clouds at large distance from the disk, to map the large-scale distribution of CRs and probe correlations with localized magnetic structures and outflows, especially in the key altitude range between a few hundred pc and a few kpc. Robust estimates of their distances, e.g., using brackets based on Gaia data \cite{gaiadr2}, would be of great help to aid the interpretation of the gamma-ray data in term of CR gradients. Furthermore, we note that observations of emission from the halo of the Milky Way may be corroborated in the near future by observations of the nearby edge-on Andromeda galaxy: claims of detection of gamma rays from its halo already exist, but for the moment it remains unclear whether it is due to CR interactions, either in the form of an escaping CR flux interacting with the intergalactic medium or of nonthermal lobes analogous to those observed in galaxies with an active nucleus, or it is of exotic nature related to hypothetical dark-matter particles annihilations \cite{pshirkov2016,karwin2019,recchia2021}.

From the theoretical point of view the last few years have shown a renewed interest in treating CR propagation in the halo on a more physical ground rather than relying on imposing a boundary condition at a height $z_{max}$ and a diffusion coefficient adjusted to reproduce the data. Several attempts have been made to explain CR diffusion as a result of the non-linear interaction with plasma waves self-excited owing to the CR streaming instability \cite{recchia2016,holguin2019,buck2020,dogiel2020}, as briefly introduced in Section \ref{sec:modda}.  
In particular, \citet{Evoli:2018nmb} discussed a scenario in which the diffusion properties of CRs are derived from a combination of wave self-generation and advection from the Galactic disc, with a halo of a few kpc naturally arising as a consequence. All of these models are non-linear, and explore the feedback from CRs on galaxy evolution via the formation of winds and the subsequent impact on star formation and galaxy evolution, while succeeding to reproduce to some extent the observed properties of primary CRs. Predictions by  \cite{Evoli:2018nmb}  are shown for illustration in Figure~\ref{fig:vertical_gradient}. Their model overpredicts CR densities at a few kpc from the disk compared to gamma-ray emissivities. However, the tuning of the model parameters did not take into account gamma-ray measurements, which serves as a nice illustration of the importance of halo emissivity measurements in this context. Further observables that can complement the halo gamma-ray emissivities are the isotropic gamma-ray background and the flux of high-energy neutrinos \cite{feldmann2013,blasi2019,recchia2021}.

\subsection{New frontiers: residual gamma-ray emission}
\label{escape_obs_res} 
One of the most interesting and surprising results from \F-LAT observations has been that extended residual emission as large as $\sim$30\% on a variety of different scales appears on top of large-scale interstellar emission from the Milky Way as accounted for via either template fitting or standard implementations of CR propagation models \citep{Ackermann:2012,Acero:2016qlg}. There are clear indications that large-scale CR trends described in the previous sections are not sufficient to capture the richness of the CR phenomenology in the Galaxy, which also mirrors the gamma-ray enhancements observed in the LMC (Section~\ref{sec:disks}).  In other words, the sensitivity and breadth of gamma-ray observations is now such that we may have reached the limits of standard basic modelling approaches.

Some of the features emerging in the residuals are considered amongst the most important results from \F.  The \F bubbles are a large bipolar structure seemingly emanating from around the Galactic center \cite{Dobler:2009xz,Su:2010qj,Fermi-LAT:2014sfa}, which has been interpreted, for instance, as the result of a hadronic wind that advects particles out of the Galactic disk, or of a leptonic jet accompanied by anisotropic diffusion along magnetic field lines that drape around the bubble surface (for a review, see, e.g., \cite{yang2018}). The Galactic center excess is a large feature with approximate spherical symmetry around the center of the Galaxy, which has attracted a lot of attention owing to the possible interpretation as the result of annihilation/decay of hypothetical dark-matter particles, but that could also find a more mundane explanation, for example in terms of unresolved populations of sub-threshold   gamma-ray sources such as millisecond pulsars (see, e.g., \cite{murgia2020} and references therein). Observations and interpretation of these features are already covered in full reviews by their own, included those just referenced, so they are left out from our paper.

Aside from speculations on possible exotic phenomena such as radiation from dark matter, three main families of solutions to the puzzle of residuals have been proposed, invoking either (\textit{i}) contributions to diffuse emission from unresolved gamma-ray sources, (\textit{ii}) undetected or poorly modelled interstellar gas and radiation fields acting as target for CR interactions, or (\textit{iii}) the effect of CR injection in the ISM localized in space and time possibly accompanied by peculiar conditions of particle transport. While in the context of this review we will be mainly concerned with the latter, let us remark that a mix of the three is the most likely explanation for the observations. We will cover in detail observations of gamma-ray emission in the vicinity of sources and their implications for CR transport in the next Section.

However, the discretized nature of CR sources may explain also large-scale residual emission. In a recent work, \citet{Porter:2019} have assessed how discrete and steady-state CR injection differs when it comes to predict gamma-ray interstellar emission. Even using simplified prescriptions for discrete injection, i.e. no localized self-confinement around sources and no spatial and temporal clustering from OB associations, the work illustrates the wealth of effects that can be expected. Compared to the corresponding steady-state case, stochastic injection can give rise to intensity features with sizes from a few to a few tens of degrees, both excesses and deficits, with amplitudes reaching 50\% and above of that predicted in the steady-state case. The effect is more pronounced at larger energies, at intermediate/high longitudes and latitudes, and for leptonic radiation processes, which constitutes interesting challenges for the new and future VHE instruments that will provide extended coverage of the Galactic plane and give access to larger angular scales (e.g., HAWC, LHAASO, or CTA). In that endeavour, one should keep in mind that simply extracting excess or residual emission in a reliable way may be challenging. \citet{Porter:2019} illustrate the biases that can result from using a reference interstellar emission model based on a mismatched steady-state smooth source distribution (e.g. including spiral arms or not). Very large scale emission structures can ensue, some of which reminiscent of the \F~bubbles. In addition, isolating excess emission from residual significance maps can lead to various troubles, such as the splitting of the true component into various substructures and a biased determination of other emission components. The most promising avenue to address such issues is to incorporate multiwavelength information and try to constrain the CR injection and propagation history from a large set of observations, from radio/microwave to X-rays and gamma rays. \label{sec:residuals}

\section{Gamma-ray emission in the vicinity of sources: the early steps of a long journey for cosmic rays}\label{sec:youngcr}
Fully understanding the life cycle of galactic CRs requires connecting the properties of particles still confined within their sources and/or undergoing acceleration to the characteristics of a galactic population built up over Myr time scales from thousands of transient accelerators of various subclasses. Recent observations highlight that the early phases in which CRs are still wandering around
 their original source play a special role within this life cycle, and have a remarkable impact on the associated non-thermal emissions. There is no unambiguous terminology for this particular moment in the CR life cycle: it is sometimes referred to as the escape of CRs, but that term may also refer to the specific process in which CRs are released by the accelerator. For that reason, and also because what happens close to the sources can be more complex than CRs escaping from a single and well-defined accelerator, and, furthermore, sources may have an impact on the older galactic CR population roaming around them, we will focus in a more generic fashion on the phenomenology of CRs in the source vicinity.

As will be developed below, there are indications that CRs are not swiftly and seamlessly transferred from their source to the ISM at large. Instead, they likely experience some confinement in the vicinity of the source, and the duration and extent of this confinement will influence their early interaction with the ISM, with possible consequences on several of the observables through which we probe the CR phenomenon: for instance, the isotopic and spectral properties of the local flux of CRs, or the morphology and spectrum of the large-scale interstellar emission \citep{DAngelo:2016,DAngelo:2018,Johannesson:2019}. Before diving into gamma-ray observations, let us mention that, although a less direct evidence, features in the local CR flux can also be interpreted as resulting from processes related to the early stages of CR propagation. This possibility is important as it competes with dark matter interpretations of anomalies in the CR signal. This is true for instance of the local positron flux, which can be influenced by the dynamics of pair release from nearby pulsars \citep{Abeysekara:2017b,Profumo:2018,Manconi:2020}. In the context of this review, however, we will leave such considerations aside.

In the context of gamma-ray observations, CRs in the vicinity of sources will give rise to emission structures lying at intermediate spatial scales between isolated objects, such as SNRs or PWNe, and the large-scale diffuse emission of the ISM. Here we will focus on how the phenomenon fits into the global gamma-ray emission of a star-forming galaxy, with particular emphasis on the Milky Way. CRs in the vicinity of sources can be associated to specific populations of gamma-ray sources or specific regions of the ISM, and we will review below the current knowledge on such objects for three categories: emission beyond the shock in SNRs, emission around pulsars and their nebulae, and extended emission coincident with star-forming regions (SFRs). 
This categorization does not necessarily corresponds to specific kinds of acceleration sites, as mixed scenarios are likely to occur: particle escape from a PWN influenced by conditions inherited from the parent SNR, or superposition of processes in rich SFRs. Before going deeper into the topic, let us mention that the topic of CRs in the vicinity of sources is still actively being explored, owing both to the complexity of the physical problem, the diversity of possible astrophysical setups, and the difficulty in giving a clear-cut interpretation to existing observations. Rapid evolutions are therefore expected in this field in the coming years.

\subsection{Physical problem}
\label{escape_phy}

The release of CRs from a source is most likely more than just a localized and temporary enhancement in CR density. First of all, in general the release of the non-thermal particle content of a source will not be instantaneous, but it will be spread over time with some energy dependence \citep{Nava:2016,Nava:2019}. Second, the energy density associated to the CR enhancement around a source will exceed the typical energy density of the ISM for long durations \citep{BlasiAmato2012}, and it will be hard to avoid some dynamical feedback of escaping CRs on the surrounding medium \citep{Schroer:2020}. Both considerations point to CR escape being a complex problem of non-linear dynamics and plasma physics, with strong dependence on both the actual history of particle release from the source and the environmental conditions around it.  

The non-thermal particle yield of an accelerator is expected to be released progressively, over time scales comparable to the lifetime of the source and in an energy-dependent way. In the specific case of diffusive shock acceleration in SNRs, the highest-energy particles in the 100~TeV-1~PeV regime detach from the accelerator in the very first few $10^2$ yr of the SNR expansion, while 1-10~GeV particles are released after $10^5$ yr, when the SNR enters the radiative stage \citep{Ptuskin:2005}. Particle escape is intimately connected to the acceleration process \citep{Drury:2011,Bell:2013}, so lacking a fully consistent and effective theory for the latter necessarily impacts our understanding of the former. That difficulty, however, can be turned into an opportunity. Since escape is so deeply rooted into the acceleration process, studying the vicinity of sources provide a complementary opportunity to address several key questions of the CR phenomenology. What is the exact CR spectrum fed into the ISM by the (different categories of) sources? What was the maximum energy attained by accelerated particles and how did it evolve in time? What fraction of the source power/energy went into CRs? What is the nature of the accelerated particles? We will illustrate below how some of these questions can be addressed in practice. Before doing so, we emphasise the relevance \citep[and challenges, see][]{Mitchell:2021} of searching for signatures of CRs around sources when trying to elucidate the question of acceleration up to the so-called knee region of the local CR spectrum, essentially because the time spent by such very-high-energy particles in the accelerator is small and the probability of finding active sources accelerating particles to PeV energies (PeVatrons) in the Galaxy is accordingly limited \citep{Cristofari:2020}.

Once decoupled from the acceleration zone, fresh CRs will influence the transport conditions in the medium surrounding the source, via the same processes that governed their confinement in the source, i.e., self-generation of magnetic turbulence from resonant and non-resonant instabilities \citep{Malkov:2013,Bell:2013}. This modifies the conditions of particle transport around the source for long durations and over large scales, with a strong energy dependence. In \citep{Nava:2019}, it is estimated that CRs escaping in a hot fully-ionized medium will experience suppressed diffusion by a factor reaching up to 20 with respect to the ISM at large within 50-100\,pc around the source and over durations of the order of several 10\,kyr for 1\,TeV particles and several 100\,kyr for 10\,GeV particles. For a medium containing neutral species, the damping of the self-generated turbulence from ion-neutral friction reduces the confinement duration by nearly a factor of 10 \citep{Nava:2016,Brahimi:2020}. In addition to this impact on the local turbulence, if the density of energy and momentum carried by escaping particles is comparable to or well in excess of what is found in the medium, major dynamical effects can result, such as the clearing of the surrounding medium by an overpressurised bubble of trapped CRs \citep{Schroer:2020}. 

These processes can be expected to give rise to a large variety of observable situations depending on the actual parameters of the problem: the stage of the process being witnessed, the particle energy being probed, the physical scales accessible to the observation, the objects and processes involved in particle acceleration, the interstellar conditions in which escape takes place and by which it is made visible to us. On the latter point, one should mention that anisotropic diffusion around sources, physically well-motivated, and resulting either from the orientation of the regular background magnetic field or from the actual topology of the large-scale turbulence modes, can produce non-trivial emission patterns hard to identify and interpret \citep{Giacinti:2013}. Indeed, as illustrated below, the observational evidence associated to escaping CRs is very diverse and its interpretation is far from unified.



\subsection{Emission beyond the shock in supernova remnants}
\label{escape_obs_snrs}

SNRs still remain the leading candidates for the acceleration of Galactic CRs, if not necessarily for the PeV-energy particles at least for the bulk of the lower-energy population, so searching around them for particles in the process of merging with the galactic population is a promising avenue, and may provide useful information complementary to that inferred from on-going acceleration. Observationally, however, the picture can be very diverse: emission from the immediate shock upstream, from particles that are detaching from the shock precursor \citep{Abdalla:2018}; emission from shocks crushing into nearby clouds, possibly causing a sudden release of the CRs trapped downstream \citep{Cui:2018} or the reacceleration of ambient CRs trapped in the clouds \citep{Uchiyama:2010,Abdollahi:2020}; emission from escaped particles that are well detached from the shock and have diffused out to some distance and illuminate large gas clouds \citep{Gabici:2007,Hanabata:2014}. In the last two cases, the overall picture can be complicated by the fact that escaping particles can diffuse back into their parent SNR, even without participating anymore in the acceleration, and contribute to the gamma-ray emission of the object \citep{Celli:2019}.

Let us focus first on emission relatively close to the shock. Deep H.E.S.S. observations of RX J1713.7-3946 illustrate the challenge of studying the early stages of CR release from an SNR, i.e., particles just detaching from the shock, even with unprecedented angular resolution allowing to probe sub-parsec scales \citep{Abdalla:2018}. Significant extension of the gamma-ray emission beyond the shock (traced by X-ray synchrotron emission) is detected but cannot clearly be attributed to the shock precursor or diffusive escape, nor can it be used to tell the exact nature of emitting particles. It is unclear whether next-generation gamma-ray instruments will provide sufficient improvement in angular resolution to revolutionize such analyses, and here it seems that multiwavelength studies will be key to advance our knowledge. On slightly larger scales, and on a slightly older object more prone to significant escape, gamma-ray emission beyond the shock of $\gamma$ Cygni was detected over a broad spectral range thanks to \F-LAT and MAGIC observations \citep{Acciari:2020}. Interpreted in a coherent framework linking acceleration and escape along the SNR's history, the observations illustrate the joint constraints that can be derived on both processes, e.g. the time evolution of the maximum CR energy, the acceleration efficiency, or the diffusion coefficient in the vicinity of the remnant. The results also point to a diffusion coefficient two orders of magnitude smaller than in the Galaxy at large.

\end{paracol}
\begin{specialtable}[htb]
\caption{List of the main established or candidate interacting SNRs detected in gamma-rays}
\begin{tabular}{cccccc}
\toprule 
Name & $\alpha$ & $\delta$ & Distance & GeV source & TeV source \\  
 & (deg) & (deg) & (kpc) & &  \\  
\midrule
SNR G006.4-00.1 (W 28) & 270.34 & -23.29 & $1.9^{+0.4}_{-0.4}$ & 4FGL J1801.3-2326e & HESS J1801-233/J1800-240A,B,C  \\
SNR G008.7-00.1 (W 30) & 271.41 & -21.61 & 4.5 & 4FGL J1805.6-2136e & \\
SNR G023.3-00.3 (W 41) & 278.72 & -08.74 & $4.2^{+0.3}_{-0.3}$ &  & HESS J1834-087 \\
SNR G034.7-00.4 (W 44) & 284.04 & 01.22 & 3.0 & 4FGL J1855.9+0121e &  \\
SNR G043.3-00.2 (W 49B) & 287.79 & 09.11 & $10^{+2}_{-2}$ & 4FGL J1911.0+0905 & HESS J1911+090 \\
SNR G049.2-00.7 (W 51) & 290.96 & 14.10 & $4.3^{+1.7}_{-0.0}$ & 4FGL J1923.2+1408e & HESS J1923+141 \\
SNR G089.0+04.7 (HBH 21) & 311.25 & +50.58 & $1.7^{+1.3}_{-1.1}$ & 4FGL J2045.2+5026e &  \\
SNR G189.1+03.0 (IC 443) & 94.51 & 22.66 &1.5 & 4FGL J0617.2+2234e & TeV J0616+225 \\
SNR G318.2+00.1 & 223.70 & -59.07 & $4.0^{+5.4}_{-0.7}$ &  & HESS J1457-593 \\
SNR G348.5+00.1 (CTB 37A) & 258.63 & -38.48 & $9.0^{+0.5}_{-2.7}$ & 4FGL J1714.4-3830 & HESS J1714-385 \\
SNR G349.7+00.2 & 259.54 & -37.31 & $11.5^{+0.7}_{-0.7}$ & 4FGL J1718.0-3726 & HESS J1718-374 \\
SNR G357.7-00.1 & 265.07 & -30.97 & $ 12.0 $ & 3FGL J1741.1-3053 &  \\
SNR G359.1-00.5 & 266.37 & -29.95 &  & 4FGL J1745.8-3028e & HESS J1745-303 \\ 
LHA 120-N132D & 81.26 & -69.64 & 50 & & TeV J0525-696 \\
\bottomrule
\label{tab_mcsnrs}
\end{tabular}
Notes: Right ascensions and declinations in degrees were recovered from the CDS and rounded to two decimals. Associations with gamma-ray sources were obtained from the Centre de Donn\'ees astronomiques de Strasbourg (CDS) and, for the TeV counterparts, we favoured the HESS naming when available and otherwise used the TeVCat naming. Distance estimates in kpc were reproduced from \citet{Acero:2016a} or from TeVCat when not available in the former reference, except for the distance to the LMC which was set to 50\,kpc. Note that LHA 120-N132D was clearly detected as a GeV source but has no specific 4FGL name.
\end{specialtable}
\begin{paracol}{2}
\switchcolumn 

Looking at larger physical scales, SNRs interacting with molecular clouds have become over the past decade a growing class of gamma-ray sources. Currently, 8 TeV sources are classified as such in the TeVCat catalog\footnote{See http://tevcat2.uchicago.edu, database version tevcat2\_test.3437}, and 11 GeV sources were classified as such in \citet{Acero:2016a}, based on coincident molecular line emission, especially OH maser emission at 1720\,MHz. The two sets overlap and we summarized the full sample in Table \ref{tab_mcsnrs} for convenience.

Although the sample is still limited and there is large scatter in the observed or inferred properties, interacting SNRs seem to be older and more luminous systems, with a softer emission spectra, compared to younger SNRs that are still in their Sedov phase and feature high-velocity shocks \citep{Acero:2016a}. In this evolutionary trend, there may be a gradual shift in emission processes, with younger systems being dominated or having a more significant contribution from IC scattering, while older interacting systems would be dominated by pion decay emission. Increased population statistics provided by the future Galactic plane survey with CTA may well reveal a less clear-cut separation between the two classes. In a few cases the data allow a (model-dependent) estimation of the diffusion coefficient around the SNR, which is found to be a factor of a few to a hundred lower than in the ISM at large \citep{Uchiyama:2012,Hanabata:2014}.

In the VHE range, more interacting SNRs are likely to be found among the $\sim$60 unidentified low-latitude TeV sources, for instance HESS J1852-000 \citep{Abdalla:2018a} or HESS J1702-420 \citep{Eagle:2020}. At intermediate energies between the GeV and TeV ranges, \citet{Eagle:2020} find a dozen unassociated low-latitude and hard-index objects in the 2FHL catalog of LAT sources emitting above 50\,GeV, and argue that some of them may be SNRs interacting with gas clouds, via direct shock interaction or escaping CRs illuminating distant gas; such an interpretation is proposed for two sources with emission coincident with the edge of an SNR \citep{Eagle:2019,Eagle:2020}. In the HE range, \citet{Acero:2016a} indicate that an additional $\sim$50 significant sources only marginally overlapping with radio SNRs are found and may be interpreted as regions of high-density gas illuminated by escaping CRs that propagated away from their source. Overall, the detected source population at GeV energies is consistent with being mostly composed of SNRs interacting with dense material with effective densities of the order of tens H~cm$^{-3}$. The numbers quoted above show the potential of CRs in the vicinity of sources to account for some fraction of the currently unidentified gamma-ray sources, both in the GeV and TeV range.

\citet{Tang:2019} investigated 10 \F-LAT SNRs\footnote{The sample of 11 sources from \citet{Acero:2016a} minus HBH 21, which is not detected above 10~GeV \cite{Ackermann:2017fges} owing to its spectral turnover at $\sim$1~GeV \cite{pivato2013,ambrogi2019}.}, testing two competing scenarios for the origin of the emission: direct interaction of the SNR shock with dense gas clouds, or escaped CRs illuminating nearby molecular clouds. The author concludes that the observed properties of the sample are inconsistent with the escape scenario, because the latter would imply a variety of spectral shapes, especially low-energy cutoffs, that is not observed. Instead, direct interaction involving reacceleration of ambient, potentially harder CRs and adiabatic compression is claimed to explain the diversity of spectral shapes of the sample, in particular the variety of high-energy breaks (as observed in W49B, W51C, or G349.7+0.2). Clearly, a consensus on the physics at play in these object is not yet reached, and it is not obvious that gamma-ray observations alone will suffice to lift the ambiguities in the interpretation.

The object W28 alone exemplifies how complex the emission scenario may be: the emission observed around the SNR is composed of 4-5 components, most of which are detected at both GeV and TeV energies, and currently proposed interpretations involve a combination of direct shock interaction, possible triggering leakage of CRs from the remnant, escaped CRs illuminating distant clouds, and the contribution of background CRs \citep{Hanabata:2014,Cui:2018}.

\subsection{Emission around pulsars and their nebulae}
\label{escape_obs_psrs}

Pulsars and their wind nebulae are highly efficient factories of non-thermal electron/positron pairs, which are produced and accelerated in the magnetosphere, the relativistic wind, and its termination shock.
PWNe are a good example of the complexity of studying particles freshly detached from their source, with a strong impact from both the original source and its surroundings, and are a remarkable constituent of the non-thermal landscape of the Milky Way, especially in the VHE range \citep{Gaensler:2006}. 

Recently, the phenomenon has acquired a new dimension with the discovery of very extended emission components beyond what was held for the boundaries of PWNe in a few systems. The pulsars concerned have ages of the order of 100\, kyr and have reached an evolutionary phase where a large fraction of the accelerated electron/positron pairs can rapidly escape from the shocked pulsar wind into the surrounding medium, for instance by leakage from a bow-shock PWNe \citep{Bucciantini:2020}, instead of being long trapped into a hot and magnetized nebula \citep[see a possible evolutionary path in][]{Giacinti:2019}. These so-called halos were originally discovered with HAWC in the TeV range around pulsars B0633+17 (Geminga) and B0656+14 \citep{Abeysekara:2017b}. The most natural interpretation for the observed signal was radiation from energetic electrons/positrons IC scattering off ambient photons.

A major result was that the intensity distribution and flux level indicates a very strong confinement of particles around the source, with diffusion being suppressed by a factor of a few hundreds compared to the average ISM value inferred from local CR measurements \citep{Abeysekara:2017b}. Halos are a growing source class that may account for a large fraction of currently unidentified extended VHE sources \citep{Linden:2017,DiMauro:2019b,Sudoh:2019,3hwc}, and that, as a population, potentially give a non-negligible contribution to the diffuse emission from the Galaxy, or at least some regions of it \citep{Hooper:2018,Linden:2018}. The phenomenon can naturally be expected to give rise to emission in other bands, and indeed the Geminga halo was later found at 10-100\,GeV energies using \F-LAT data \citep{DiMauro:2019a}. A broadband picture of pulsar halos is however largely missing today and searches are currently aiming at uncovering or expanding the population of halos in the radio, X-ray, and GeV bands.

Pulsar halos constitute a good opportunity to study particle transport around sources, especially because the radiating pairs are energetically subdominant in the medium \citep{Giacinti:2020}. Self-confinement by the streaming pairs could be responsible for diffusion suppression in the early phases, but the challenge in the case of middle-aged pulsars like Geminga is to sustain that confinement at later times, when the pulsar spin-down power has much decreased \citep{Evoli:2018}. An alternative explanation is the presence of fluid turbulence injected at small scales to guarantee sufficient power at the scales relevant for 100\,TeV particle scattering \citep{LopezCoto:2018}. Such conditions could be inherited from the expansion of the parent SNR. Another possibility is that escaping pairs are experiencing the turbulence imprinted in the vicinity of the system by CRs escaping from the parent SNR, although here again the question of maintaining such a turbulence over several 100\,kyr should be further investigated. In any case, pulsar halos offer a great opportunity to study CR transport in the vicinity of some accelerators.

\subsection{Emission coincident with star-forming regions}
\label{escape_obs_sfrs} 

SFRs, especially the most extensive ones involving massive stars, are expected to be prominent objects in the gamma-ray sky. First, because a large fraction of the most promising sites for particle acceleration will be found clustered in SFRs (colliding-wind binaries, pulsars and their nebulae, SNRs). Second, because SFRs are rich in targets for CR interactions, massive gas remainders from the parent molecular clouds and radiation fields enhanced by the many luminous stars, which guarantees an efficient conversion of CR energy into gamma rays. Therefore, SFRs seem to offer optimal conditions when it comes to studying how CRs are transferred from accelerators to the galactic population, by providing frequent injection of accelerated particles and favourable conditions to observe them. 

The reverse side of the medal, however, is that the clustering of high-energy objects in the same region of space and time makes it difficult to unambiguously interpret gamma-ray observations of limited angular resolution and clearly associate a gamma-ray source with particles released from a specific object. From stellar evolution data used in \citep{Ferrand:2010}, supernova explosions will be nearly uniformly distributed in time between 3\,Myr after the initial burst of star formation (for the most massive 120\,M$_{\odot}$ stars) and 37\,Myr (for the least massive 8\,M$_{\odot}$ stars). As soon as the SFR hosts more than a few hundreds massive stars, the average time interval between supernova explosions is less than the gamma-ray lifetime of many high-energy sources ($\sim$50\,kyr for SNRs or PWNe and $\sim1$\,Myr for pulsars and their halos or colliding-wind binaries), and so a superposition of gamma-ray-emitting objects can be expected.

Furthermore, it has long been speculated that the clustering of high-energy objects may play a distinctive role in the origin of CRs \citep{Montmerle:1979,Bykov:2001,Parizot:2004}. This may occur through a variety of possible processes, for instance long-lived particle acceleration at the termination shocks of stellar winds \citep{Seo:2018,Morlino:2021}, repeated shock acceleration or turbulent reacceleration in the interior of superbubbles and/or at their bounding shell \citep{Ferrand:2010,Tolksdorf:2019,Bykov:2020}, acceleration at the highest energies in converging shock flows \citep{Bykov:2018}. Although solid observational evidence of this class of phenomena is still largely missing, a link between CR origin and SFRs is supported by the isotopic composition of CRs \cite{binns2016,murphy2016,israel2018}. Therefore, the potential of SFRs to bring answers to key issues in CR astrophysics justifies a continued search to detect and characterize their gamma-ray emission. Yet, as illustrated below, isolating that specific contribution amidst a variety of concurrent particle accelerators and gamma-ray emitting sources is a real challenge.

\begin{specialtable}[H]
\caption{List of the most prominent star-forming regions with established or potential detections in gamma-rays}
\begin{tabular}{cccccc}
\toprule 
Name & $\alpha$ & $\delta$ & Distance & GeV sources & TeV sources \\  
 & (deg) & (deg) & (kpc) & &  \\  
\midrule
Cygnus & 307.17 & 41.17 & 1.3-1.8 & 4FGL J2028.6+4110e & TeV J2031+406 \\
GC & 266.42 & -29.01 & 8.5 & & TeV J1745-290 \\
 &  &  &  &  & TeV J1745-290d \\
NGC~3603 & 168.83 & -61.26 & 6-8 & 4FGL J1115.1-6118 &  \\
Westerlund 1 &  251.77 & -45.85  & 4-5 & 4FGL J1645.8-4533 & HESS J1646-458 \\
 &  &  &  & 4FGL J1648.4-4611 & \\
 &  &  &  & 4FGL J1649.2-4513 & \\
 &  &  &  & 4FGL J1650.3-4600 & \\
 &  &  &  & 4FGL J1652.2-4516 & \\
Westerlund 2 & 155.99 & -57.76  & 4-6 & 4FGL J1023.3-5747e & HESS J1023-575 \\
W~43 & 281.88 & -01.94  & 5-7 & 4FGL J1847.2-0141 & HESS J1848-018 \\
 &  &  &  & 4FGL J1848.6-0202 & \\
 &  &  &  & 4FGL J1848.7-0129 & \\
30 Doradus C & 83.96 & -69.21 & 50 &  & 30 Doradus C \\
\bottomrule
\label{tab_sfrs}
\end{tabular}
Notes: Right ascensions and declinations in degrees were recovered from the CDS and rounded to two decimals. Associations with gamma-ray sources were obtained from the CDS, complemented by the 4FGL \cite{Abdollahi:2020} and the H.E.S.S. Galactic Plane Survey \cite{hgps} catalogs. Some SFRs are associated to multiple 4FGL pointlike sources when no dedicated extended template was included in the automated catalog analysis. For the TeV counterparts, we favoured the H.E.S.S. naming when available and appropriate, and otherwise used the TeVCat naming. Typical distance estimates come from determinations and/or literature review in \cite{Berlanas:2019} for Cygnus, in \cite{Melena:2008} for NGC~3603, in \cite{Beasor:2021} for Westerlund 1, in \cite{Furukawa:2009,Vargas-Alvarez:2013} for Westerlund 2, and in \cite{Nguyen-Luong:2011} for W~43. Note that the ranges of distances result from both the difficulty in accurately locating the region in the Milky Way and its actual spread along the line of sight (e.g. substructures, in the Cygnus OB2 association). Typical distances of 8.5 and 50\,kpc are used for the GC and LMC, respectively.
\end{specialtable}

As of today, GeV emission has been detected in the direction of about half a dozen SFRs. The most prominent SFRs studied in our Galaxy are listed in Table \ref{tab_sfrs}: the Cygnus region \cite{Ackermann:2011}, NGC~3603 \cite{Saha:2020}, Westerlund 1 \cite{Ohm:2013}, Westerlund 2 \cite{Yang:2018}, and W~43 \cite{Yang:2020}. Several of these targets are also associated to emission in the TeV range, for instance Cygnus \cite{Bartoli:2014,HWCcocoon2021}, Westerlund 1 \cite{Abramowski:2012}, or Westerlund 2 \cite{Abramowski:2011}. We can also include in this category the Galactic center, in which diffuse TeV emission is detected in the direction of three super-massive star clusters, although the CR source there may alternatively be connected to the central black hole \citep{Abramowski:2016}. Outside of the Milky Way, a handful of SFRs may be studied with existing gamma-ray instruments, for instance 30 Doradus or the N11 region in the LMC, NGC~602 and NGC~346 in the SMC. The 30 Doradus region was observed at both GeV and TeV energies \citep{Abramowski:2015,Ackermann:2016}. The entire region is detected at GeV energies only, although with no apparent specific feature, while only a more peripheral emission was detected at TeV energies only in the direction of the 30~Doradus~C superbubble.

Over recent years, the list has been rapidly expanding thanks to more candidates emerging in gamma-ray source catalogs and a growing number of dedicated studies. The latest 4FGL-DR2 revision of the main \F-LAT catalog contains five associations of gamma-ray sources with SFRs: beside Cygnus and Westerlund~2 that were already mentioned, we find $\rho$~Ophiuchi and the \hii region Sh~2-152 in our Galaxy, and NGC~346, which is the brightest star-forming region in the SMC  \cite{Ballet:2020}. The general LAT catalogs are mainly aimed at pointlike sources, therefore important complementary information is gathered by catalogs targeting extended sources, which revealed emission in the direction of W~30 \cite{Ackermann:2017fges}, and of NGC~7822, NGC~1579, and IC~1396 \cite{ackermann2018fhes}.  The latter work, however, illustrates the difficulties of disentangling the potential emission of a SFR from the foreground and background emission in a catalog-like analysis.

Among dedicated studies investigating the link between gamma-ray emission and SFRs we mention for instance those of W~30 \cite{Liu:2019w30}, W~40 \cite{Sun:2020}, or the \hii region G41.1-0.2 \citep{Ergin:2021}. In the direction of region G25.0+0.0, extended gamma-ray emission was detected and, based on similarities with Cygnus and a positional correlation with gas structures and energetic sources, it was proposed to be associated with a putative candidate OB association \cite{Katsuta:2017}. 

A common feature of several of these gamma-ray sources is their extended morphology and relatively hard spectrum, which is exactly what one would expect from young CRs freshly detached from their sources and diffusing away into the ISM. Where both GeV and TeV detections are available, however, it is frequent to observe mismatch and offsets in the GeV and TeV morphologies \citep[see e.g.][for Westerlund 1]{Ohm:2013}. This is often explained in terms of superposition of sources, as anticipated above \citep[e.g. an SNR interacting with a molecular cloud and a pulsar/PWN system in the case of W30; see][]{Liu:2019w30}. To date, the most convincing association between gamma-ray source and SFR remains Cygnus, where the morphology of the gamma-ray emission at GeV energies shows a striking resemblance with the cavities carved in the ISM by the stellar winds and ionization fronts \cite{Ackermann:2011}.  

For some of these sources, the gamma-ray data were used to infer a radial distribution of CRs around the presumed accelerator and derive constraints jointly on the CR injection history and transport properties. A $1/r$ radial profile determined for a handful of SFRs was recently invoked as evidence that SFRs are continuously releasing CRs over several Myr, presumably from the conversion of stellar wind power into CRs with 1-10\% efficiencies \cite{Aharonian:2019}. For one of these regions, Westerlund~2, the diffusion coefficient was estimated to be 100 times smaller than in the large-scale ISM. While the radial CR distribution is indeed a powerful discriminant to characterize the acceleration site and transport process, establishing it from the gamma-ray observations is a highly non-trivial task.

First, the extended gamma-ray emission from the region of interest along the line of sight needs to be separated from the foreground and background emission in the Galaxy. Second, the conversion of gamma-ray intensity profile into CR density profile requires a robust knowledge of the gas mass distribution in/around the SFR. Unfortunately, both steps can be shaky. The first point was recently illustrated in the case of NGC~3603, where an updated Galactic interstellar emission model and a better treatment of other extended sources in the field resulted in the source associated with NGC~3603 not being significantly extended \citep{Saha:2020}, contrary to earlier claims \citep{Yang:2018}, even without resorting to dedicated interstellar emission models such as those developed for an analysis of the Cygnus region \citep{Ackermann:2011}. As to the second point, the lack of apparent correlation of the gamma-ray emission with the ambient gas distribution, e.g. for Westerlund 1 and 2 or Cygnus, even accounting for some radial CR distribution, should at the very least call for caution in the interpretation of the observations. In that respect, the $\sim$1-10~TeV emission from the Galactic center and its correlation with gas in the central molecular zone may still constitute the most convincing evidence for a $1/r$ radial CR density profile \citep{Abramowski:2016}. It is interesting that even in such conditions, the identification of the CR source remains elusive. The CR gradient at the center of the galaxy can be equally well explained from continuous CR injection by a central stationary source such as Sgr A* active over several Myr \citep{Abramowski:2016}, or from stochastic CR injection by dozens of SNRs over 100\,kyr \citep{Jouvin:2017,Jouvin:2020}.

In an effort to identify the specific role of SFRs in the CR lifecycle, it is equally interesting to consider those objects that were not detected yet. For instance, \F-LAT observations of eight young star clusters with ages below a few Myr constrain the particle acceleration efficiency in stellar winds to be below 10\% in four of them and below 1\% in 2 of them \citep{Maurin:2016}. For more evolved objects, it is striking that the extraordinary 30 Doradus region in the LMC does not stand out in GeV gamma-rays: after removing the emission from two very powerful pulsars lying in the field, the region seems to fit perfectly in the larger scale-emission of the galaxy and does not stand out in gamma-rays, or at least not in proportion of its ionizing luminosity \citep{Ackermann:2015LMCpsr,Ackermann:2016}. Looking at even more evolved structures, it was recently established that the Orion-Eridanus superbubble does not harbour any significantly enhanced CR population at GeV energies with respect to the solar neighborhood \cite{joubaud2020}. It is not clear yet if this has to be attributed to time evolution, stellar content, or its less compact nature compared to, e.g., Cygnus, but this non detection needs to be taken into account when assessing the potential of superbubbles for turbulent acceleration/reacceleration \citep{Tolksdorf:2019,joubaud2020}.

\section{The population of gamma-ray emitting galaxies: different realizations of the cosmic-ray phenomenon}\label{sec:galaxies}
In this section we will turn our attention to the population of external galaxies whose gamma-ray emission is not dominated by the activity from a central supermassive black hole. We will frequently refer to these as star-forming or starburst galaxies (hereafter SFGs and SBGs), although we recognize that such a naming is not fully appropriate because there are numerous examples of active galactic nuclei accompanied by star formation (establishing the relationship between the two phenomena being an active field of research).

Going beyond the three external galaxies spatially resolved by current gamma-ray telescopes (covered in Section~\ref{sec:largescale}), the interest of studying the integrated emission from gamma-ray emitting galaxies as a population is at least threefold:
\begin{enumerate}
\item it allows an investigation of the physics behind CR acceleration and transport in a wide variety of conditions that differ markedly from those typical of the Milky Way (ISM gas structure and densities, radiation and magnetic field intensities, role of strong galactic winds,...), thus providing a good test bed for our current understanding of CR physics;
\item it is needed to pinpoint the actual integrated contribution of star-forming galaxies to various extragalactic backgrounds, especially the gamma-ray background detected in the energy interval 0.1-820\,GeV with the \F-LAT, and the diffuse astrophysical neutrino background observed in the energy interval 10\,TeV-10\,PeV with IceCube;
\item it offers a probe of feedback processes in galaxy evolution, because CRs can play an active role in mediating this feedback (e.g. by driving gas-loaded winds and outflows off the galactic disk, thus removing material for star formation, or by ionizing molecular gas, hence altering the conditions for it to take place), or simply because CR interactions can provide information on the ongoing processes affecting the matter cycle in galaxies.
\end{enumerate}

In the following, we focus on the first two items. We first review the current status of gamma-ray observations before addressing how they can improve our understanding of CRs.
Concerning the last item, an example for recent progress in incorporating gamma-ray observations in galaxy evolution works can be found in \citep{Chan:2019}. 

\subsection{A growing source class: the gamma-infrared luminosity correlation and its implications}
\label{galaxies_obs}

SFGs and SBGs were foreseen as a distinct GeV to TeV gamma-ray source class well before their actual detection \citep[see e.g.][]{Voelk:1996,Torres:2004}. As of today, gamma-ray emission has been established from a dozen SFGs and SBGs, mostly at GeV energies using the \F-LAT, and among these only two starbursts have also been observed at TeV energies. In a recent systematic search for gamma-ray emission from a sample of 588 SFGs using \F-LAT, 11 objects were firmly detected in the $\sim$0.1-100\,GeV range \citep{Ajello:2020}. For 2 out of the 11 (NGC~4945 and NGC~1068) and 2 additional candidates (NGC~2403 and NGC~3424), the detected emission cannot be dominantly attributed to processes linked to star formation, and might be contaminated by an active galactic nucleus \citep{Ajello:2020,Yoast-Hull:2014,Ackermann:2012}. Among the 11 galaxies, M~82 and NGC~253 were also detected at TeV energies with the current generation of IACTs \citep{Acero:2009,Abdalla:2018d}.

Although still limited, the detected sample covers a variety of galactic properties, ranging from dwarfs (SMC and LMC) to large spirals (M31), and from relatively quiescent to starbursting objects (Arp~220). Investigating how the inferred total gamma-ray luminosity for detected and non-detected objects scales with global galactic properties, evidence was found for a correlation of the former with the star-formation rate, as traced for instance by the infrared luminosity (in the 8-1000\,$\upmu$m band). Initially investigated in \citep{Abdo:2010b} and \citep{Ackermann:2012}, the correlation was recently revisited in \citep{Ajello:2020} and is now more firmly established with a significance close to 5$\sigma$. The relation between the two quantities appears mildly non-linear, with gamma-ray luminosity evolving as infrared luminosity to the power $\sim$1.3. In fitting the observed correlation to a power-law, however, a large dispersion is obtained and it is still not clear whether that is intrinsic, or due to biases and uncertainties in the galactic parameters adopted (e.g. distances or infrared luminosities), or a combination of both effects. The gamma-ray to infrared luminosity correlation is actually reminiscent of the far-infrared to radio correlation, both in their form and commonly accepted physical explanation, and several authors actually investigated them in a unified approach \citep{Lacki:2010,Martin:2014}.

The correlation is thought to be driven by massive-star formation, which is at the origin of strong UV/optical light and CRs that, upon interaction with the ISM, power infrared and gamma-ray emission, respectively. The commonly accepted interpretation of the non-linear relation between gamma-ray emission and star-formation rate (traced by infrared) is that of an increasing calorimetric efficiency of the galaxies with respect to CRs. The calorimetric efficiency is defined as the fraction of CR power that is deposited in the ISM relative to that initially injected. High star-formation rates are reached in galaxies harbouring large amounts of dense molecular gas packed in small regions, typically $\sim$$1-10 \times 10^8$\,M$_\odot$ in a volume spanning a few hundred pc. The average gas volume density in such regions can reach a few 1000\,H\,cm$^{-3}$ and, due to the high density of young stars, the interstellar radiation field energy densities can exceed a few 1000\,eV\,cm$^{-3}$ (for comparison, typical values for the Milky Way are 1\,H\,cm$^{-3}$ and 1\,eV\,cm$^{-3}$). As a consequence, CRs lose their energy much more efficiently, in particular through radiative processes such as nucleon-nucleon inelastic collisions and IC scattering. The Milky Way is thought to be a poor proton calorimeter but a good electron calorimeter, with efficiencies of of the order of 1-2\% and 40-80\% respectively, depending on the transport scenario assumed \cite{Strong:2010} (these efficiencies were computed for radiative processes only; in terms of emission the low calorimetric efficiency of protons is compensated by the fact that they are a factor of a hundred more numerous than electrons in CRs). In contrast, a starburst galaxy like Arp~220 can reach a calorimetric efficiency above 80\% for protons in some propagation models \citep{Krumholz:2020}. The argument for increasing calorimetry in high star-formation rate galaxies seems backed up by the fact that their gamma-ray spectrum is observed to be hard, or at least harder than in the Milky Way, with a photon index $\sim$2.2-2.3 \citep{Abdalla:2018d,Ajello:2020}, which is expected if the emission is mostly hadronic in origin and CR transport is loss-dominated.

Although such a scheme is very reasonable as a general trend, there are several caveats in the interpretation of the still limited sample of galaxies available today. First of all, there is not yet a clear consensus on how CR transport evolves or differs among galaxies, and this inevitably affects our ability to predict the corresponding gamma-ray emission. For example, advection-dominated transport in high-star formation rate galaxies with powerful winds could provide an alternative explanation to the hard spectra. This will be discussed in more detail in the next section. Furthermore, most galaxies in the sample are detected as point-like objects at GeV energies, and as such the gamma-ray emission is a blend of various sources: interstellar emission, populations of sources such as SNRs or young and recycled pulsars, etc. In that respect, the only spatially resolved galaxies so far show diverse and rather unexpected pictures. The distribution and flux levels of the emission from the LMC, SMC, and M~31, as summarized in Sect. \ref{sec:disks}, call for caution in interpreting the existing observations of the population of SFGs and SBGs, especially at low star formation rates. Continuous efforts are needed to extend the sample of gamma-ray-emitting SFGs, both in number and spectral coverage.

In that respect, extending the spectra towards the MeV and TeV ranges would certainly be instrumental in securing our understanding of the emission from SFGs and SBGs. On one hand, observations of emission in the hard X-ray/soft gamma-ray bands, with NuSTAR of future instruments like ASTROGAM/AMEGO/GECCO, can probe the population of secondary and tertiary leptons in SFGs and SBGs, which is an indirect diagnostic of the degree of calorimetric efficiency in individual systems. On the other hand, in the VHE range, more detections of galaxies are crucially needed to enlarge the sample beyond M~82 and NGC~253 and allow for a deeper broadband population study. The higher sensitivity observations $>100$\,GeV with CTA and $>1$\,TeV with HAWC, LHAASO, and, possibly, SWGO, are expected to provide an extended spectral coverage relevant to study all those effects that are more likely to show up at $>$TeV energies in SBGs: energy-dependent escape, photon-photon absorption, emission from unresolved population of sources such as PWNe and pulsar halos. Also incorporating 100\,MHz-10\,GHz radio observations, with their much better angular resolution, in the interpretation of gamma-ray observations of SFGs and SBGs can help to separate emission components and add useful information, e.g., on the intensity distribution in the galactic wind, for those systems that are viewed mostly edge-on \citep{Heesen:2009}.

Beyond individual detections towards infrared-selected targets, SFGs and SBGs are expected to contribute to the extragalactic gamma-ray emission as an unresolved component, integrated over cosmological distances. Because their emission results from hadronic interactions (only partially for moderately star-forming galaxies like the Milky Way, but predominantly for starbursts like M~82 or Arp~220), they are also expected to contribute some background emission of neutrinos. Over recent years, advances in the understanding of the composition of the extragalactic gamma-sky above 50\,GeV have seriously constrained the possible contribution from SFGs and SBGs. From improved catalogs of sources and photon counting statistics, the extragalactic source population is now constrained to be composed at 86\% of blazars, thus leaving SFGs and SBGs as subdominant component \citep{Ackermann:2016b}. A forward modelling of that contribution based on luminosity functions for star-forming galaxies and the observed gamma-ray to infrared luminosity correlation yield a contribution of the order of 10\% of the extragalactic gamma-ray background \citep{Bechtol:2017,Sudoh:2018,Ajello:2020}, and a corresponding contribution at the 1\% level to the 10\,TeV-10\,PeV astrophysical neutrino flux detected with IceCube \citep{Aartsen:2015}. It should be noted, however, that the latter constraint relies on the very strong assumption that the CR/gamma-ray properties inferred in the \F-LAT energy range can be readily extrapolated up to $>10$\,PeV. Here, more gamma-ray observations at the highest energies with LHAASO, HAWC, CTA, and SWGO are crucially needed. In addition, a more realistic modeling of SBGs, taking into account a diversity of spectra in the population, as actually observed in \citep{Ajello:2020}, allows a larger contribution to the neutrino background, while still being consistent with the extragalactic gamma-ray background \citep{Ambrosone:2021}.

\subsection{How do cosmic-ray transport properties vary with galactic environment?}
\label{galaxies_mod}

The sample of detected SFGs and SBGs spans 3-4 orders of magnitude in star formation rate, average gas density, or interstellar radiation field intensity \citep[see][for some examples of extreme values]{Peretti:2019}. This set of very diverse conditions can be expected to give rise to markedly different CR populations, in particular as a result of CR transport in the ISM being very unlike that prevailing in the Milky Way (so far, to our knowledge, a more limited attention was paid to whether/how extreme galactic conditions could alter the population of accelerated particles fed by sources into the ISM). As such, the study of SFGs and SBGs in gamma rays is a very good test bed for CR physics. After about a decade since the very first detections of external SFGs and SBGs in gamma rays (beyond the LMC, already observed with EGRET), however, it is fair to say that this new class of gamma-ray sources has not dramatically modified our understanding of CRs in galaxies and there is still room for deeper studies. 
Below, we review our present ideas on how CRs may evolve in galaxies different from the Milky Way.

The efforts in modelling or interpreting the observed population can be broadly separated into two kinds: those aiming at reproducing the spectra of one or several detected individual objects, and those concerned with reproducing the gamma-ray to infrared luminosity correlation. In both cases, as already mentioned, CR injection into the ISM was treated with prescriptions similar to those used for the Milky Way, and it is rather on the side of CR transport that alternative scenarios were explored. The main processes regulating CR transport in the ISM were assessed in the specific context of SFGs and especially SBGs. Examples of relevant questions are: 
\begin{itemize}
\item Energy losses: what are the properties of the various ISM components with which CRs interact (gas, radiation and magnetic fields)? What is the actual gas structure in terms of gas phase and relative volume filling factor? Do CRs experience large-scale volume-averaged ISM conditions or are they confined to specific phases?  
\item Diffusion: what is the nature of externally-driven turbulence in extreme environments such as the cores of SBGs (strength, injection and cutoff scales)? How will the self-generation of turbulence by CRs proceed in potentially very dense and weakly ionized media? If CR transport can be approximated by diffusion, what is its the form of the diffusion coefficient, its normalization and momentum dependence?
\item Advection: what is the exact role of advection in CR transport in those galaxies where large-scale winds are observed or expected? What is the structure of the wind and how is it connected to the galactic disk? In particular, is it fast and distributed enough at its base to significantly affect CRs in the disk?
\end{itemize}
In the following, we illustrate how some of these questions were addressed in recent works and with what conclusions. We note that having at our disposal an open-source, versatile model, analogous to GALPROP, DRAGON, or PICARD (see Sect. \ref{sec:modda}), that can be easily configured and adapted to different galaxies would be ideal to go beyond simple leaky-box models and allow easier comparison of various transport scenarios. 

The extreme conditions encountered in some SFGs, and especially in SBGs, translate into much stronger energy losses than those experienced by CRs in the average ISM of the Milky Way. This effect alone has already significant implications. In \citep{Martin:2014}, CR transport is assumed to be very similar to that in the Milky Way (e.g. same diffusion coefficient), and the author assessed how the global interstellar gamma-ray emission is influenced by interstellar conditions, gas mass and distribution and the related radiation and magnetic fields, for molecular gas densities increasing from 5 to 500\,H$_2$\,cm$^{-3}$ (which is a bit short to describe extreme SBGs like Arp~220). The impact is evaluated in three different energy ranges, 0.1-10\,MeV, 0.1-100\,GeV, and 0.1-100\,TeV. As gas density increases (and so do star formation and radiation and magnetic fields), the MeV range sees a strong increase of IC and Bremsstrahlung emission and dominance of secondary electrons and positrons as radiating particles. In the GeV range, emission from nucleon-nucleon inelastic collisions is the dominant process, but its intensity increases slower than linearly with gas density and flattens as CR protons and nuclei propagation becomes loss dominated, which yields a harder spectrum with a 2.3-2.4 photon index. In the TeV range, while emission from IC scattering from primary electrons can be dominant or comparable to that from nucleon-nucleon inelastic collisions for the lowest average densities, the latter overwhelms the former at large densities because it increases almost linearly with density, as CR protons and nuclei transport is largely diffusion-dominated at the highest energies. 

One effect not investigated in \citep{Martin:2014} is that of photon-photon opacity in the densest radiation fields typical of SBGs. This was included in works such as \citep{Peretti:2019} or \citep{Yoast-Hull:2015}, and the most interesting effects are the softening of the emission spectrum beyond TeV energies, and the generation of an additional population of CR electrons at TeV energies that, in the case of hard enough injection and strongly inhibited diffusion, yields a dominant IC contribution in hard X-rays. Both \citep{Martin:2014} and \citep{Peretti:2019} (see also \citep{Lacki:2013}) illustrate the potential of the hard X-ray/soft gamma-ray range as a diagnostic of CR interactions in dense galaxies, in particular for an indirect evaluation of the calorimetric efficiency of the system. The works of \citep{Peretti:2019} and \citep{Wik:2014} show that improving current hard X-ray constraints from NuSTAR by a factor of a few may become interesting (in the case of NGC~253).

The above statements are dependent on the respective diffusion and advection properties assumed, and it actually remains to be clarified how both processes evolve among different galaxies and in which regime one or the other may dominate CR transport. Works like \citep{Peretti:2019} or \citep{Yoast-Hull:2013} succeeded in reproducing the gamma-ray emission spectrum from selected SBGs with a model in which wind advection is the dominant spatial transport mechanism (either by neglecting diffusion or using a small coefficient). In contrast, the model of \citep{Krumholz:2020} uses dedicated diffusion schemes but no advection, and is able to achieve the same goal. At the population level, the observed gamma-ray to infrared luminosity correlation can be fairly well reproduced with a Milky-Way-like diffusion scheme \citep{Martin:2014}, but an advection scheme with star formation dependent wind velocity may be a viable alternative solution \citep{Kornecki:2020}. 

\begin{figure}[htb]
\includegraphics[width=0.7\textwidth]{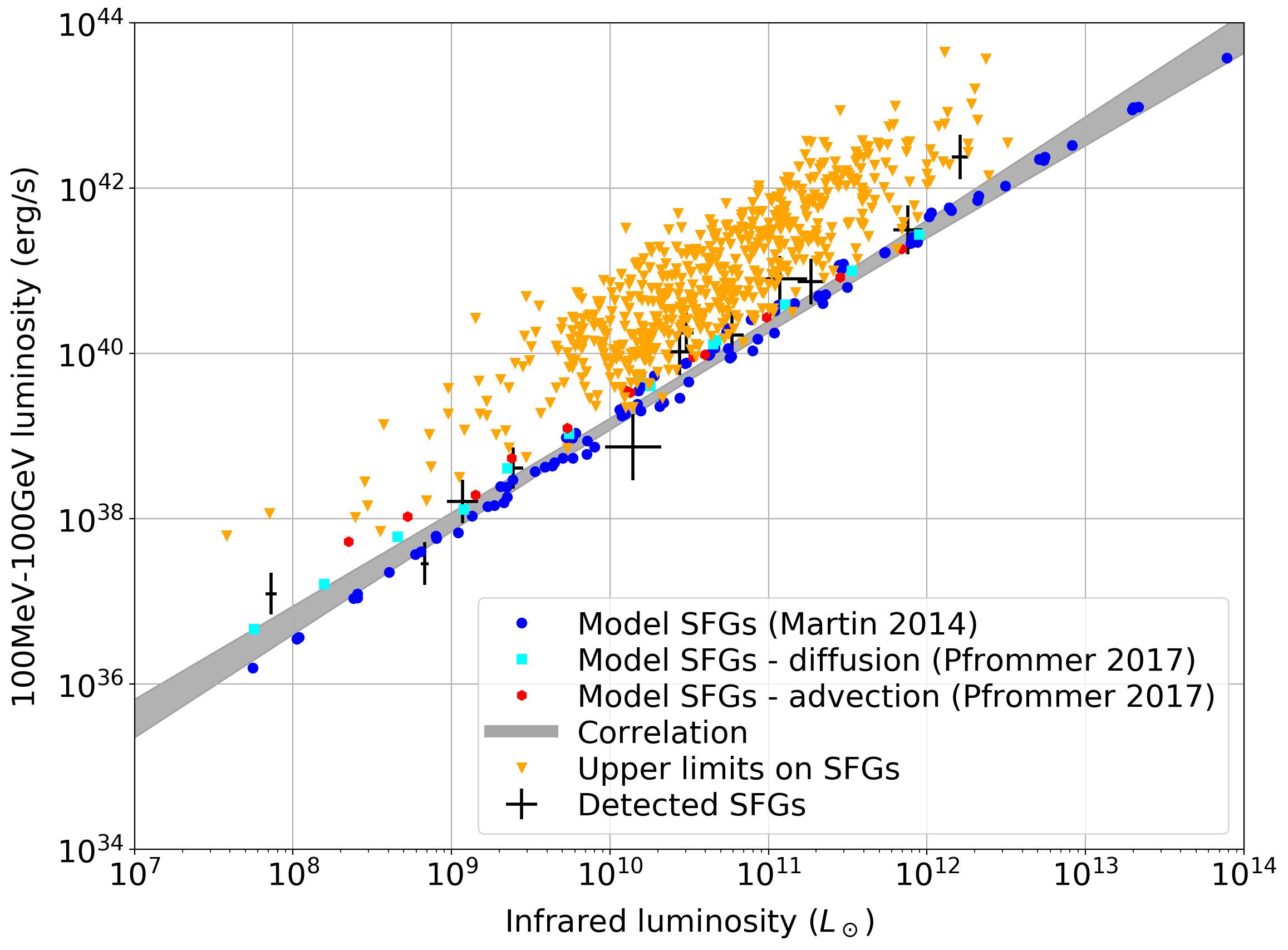}
\caption{Gamma-ray to infrared luminosity correlation. The observed luminosities plotted as black crosses and orange triangles come from \citet{Ajello:2020} and were converted to the 0.1-100\,GeV range under the assumption that the emission spectrum is a power law spectrum with photon index 2.2 (the average value measured by the authors). The correlation band plotted in gray is the one corresponding to the so-called combined fit. Overlaid are predicted luminosities: from \citep{Martin:2014}, as blue dots, for a GALPROP-like model of galaxies with different size and gas distributions and under the assumption of a Milky-Way-like CR diffusion scheme; from \citet{Pfrommer:2017}, as red hexagons and cyan squares, for a series of two-fluid magneto-hydrodynamical galaxy formation simulations for a variety of dark matter halos and under two different assumptions for CR transport: pure advection, and advection plus anisotropic diffusion (see text).}
\label{fig:gev-correlation}
\end{figure} 

Figure \ref{fig:gev-correlation} shows the most recent published version of the gamma-ray to infrared luminosity correlation, based on data from \citet{Ajello:2020}. Observations are compared to predictions from \citep{Martin:2014}, based on a GALPROP-like model, and from \citet{Pfrommer:2017}, based on an alternative modelling approach. In the latter work, the authors evaluated the gamma-ray emission in a series of two-fluid MHD galaxy formation simulations for a variety of dark matter halo masses and distributions. CRs are injected as a relativistic fluid following SN explosions and two CR transport scheme are considered: pure advection with gas motions, and advection plus anisotropic diffusion along magnetic field lines (with a single energy-independent value for the diffusion coefficient). Such an approach allows to track self-consistently and in a time-dependent way the interplay between CRs, gas motions and outflows, and galactic magnetic field structure, but it comes at the expense of approximations on the CR physics. Nevertheless, the authors show that, if a typical 10\% of the supernovae kinetic energy goes into CRs, the observed gamma-ray to infrared luminosity correlation can be fairly well reproduced and argue that it is little dependent on uncertainties in the CR transport. At high star formation rates, most of the CR energy is lost to hadronic interactions, in agreement with the calorimetry argument put forward in other works; at low star-formation rates, however, the model overpredicts the emission and the authors suggest that more realistic multi-phase ISM descriptions would be needed in that range. Such works support a trend towards numerical models in which the complexity of CR physics is dynamically coupled to the galactic environment, models whose interest reaches far beyond high-energy astrophysics (e.g. cosmological simulations of galaxy formation).

In exploring CR spatial transport in external galaxies, the work of \citep{Krumholz:2020} offers an interesting alternative to the generic prescriptions mainly in use so far. The authors evaluate that starburst regions will be dominated by volume-filling, cold and weakly ionized gas, unlike the Milky Way, where the ISM is mostly filled with hot and warm ionized gas. This implies an efficient damping of MHD turbulence modes, from ion-neutral friction, and a cutoff of the externally-driven turbulence cascade at scales such that particles with energies below a few hundreds TeV are not efficiently scattered. The streaming of CRs is then controlled by self-generated turbulence only, which is demonstrated to occur at the Alfv\'en speed for particles $\leq1$ TeV and to increase with some power $>1$ of energy above. The corresponding macroscopic diffusion scheme implies a specific flavour of energy-dependent diffusive escape such that the resulting hadronic-dominated gamma-ray spectrum follows the CR injection spectrum up to $\sim$100\,GeV and gradually falls off at higher energies (also because photon-photon absorption comes into play). This is shown to provide a convincing description of SBGs M~82, NGC~253, and Arp~220. In that picture, advection in a galactic wind is not needed but not excluded either. The authors argue that, while most CRs should be injected into the cold gas phase to ensure a high calorimetric efficiency and a sufficient level of gamma-ray emission, some CRs can be advected away from the plane with the hot ionised phase but will contribute few gamma rays (and retain the hard injection spectrum anyway).

\end{paracol}
\begin{paracol}{2}\switchcolumn
\section{Summary and perspectives}\label{sec:conclusions}
In the past decade gamma-ray observations have shed a new light on the richness and diversity of the processes that govern the build-up of CR populations in galaxies. While there is an overall agreement with the foundations of the standard Galactic CR paradigm, gamma-ray data, along with direct CR measurements and other related observables, have highlighted the limits of some standard assumptions and modelling approaches, and, therefore, have spurred many new developments in the field. We summarize below the main points highlighted in this review, before discussing forthcoming prospects and challenges.
\begin{itemize}
\item Gamma-ray emission from the local interstellar medium (Sec.~\ref{sec:locemiss}) as seen by the \F~LAT shows an overall consistency with direct CR measurements. However, a few hints of deviations are arising: fluctuations $\lesssim 25\%$ in the emissivity of atomic hydrogen between different regions within 1~kpc from the Sun; an excess of $\sim$20-30\% in the average emissivity of atomic hydrogen above 10 GeV; possible spectral deviations in a few molecular clouds. In order to definitely establish whether those deviations are related to CR injection and transport or to observational biases we need to improve our understanding of interstellar medium tracers and hadronic gamma-ray production cross sections. Complementary results are expected in the near future in the TeV energy range.
\item The large-scale distribution of CR nuclei throughout the Milky Way disk (Sec.~\ref{sec:diskmw}) as inferred from \F-LAT data shows a decline milder than expected toward the outer Galaxy and an increase by a factor of 1.5-3 in the inner Galaxy accompanied by a spectral hardening. Although alternative explanations, such as a contribution larger than currently believed of unresolved gamma-ray sources to diffuse emission, are still not completely ruled out, these trends may have deep implications for the physics of particle transport in the Galaxy (Sec.~\ref{sec:gradhard}). Competing explanations include non-homogeneous diffusion related to turbulence generation around CR accelerators,  the difference scaling with energy of parallel and perpendicular diffusion coupled with the peculiar structure of the Galactic magnetic field, and non-linear transport due to CR-driven instabilities. With the first detections of Galactic diffuse emission by current experiments, very-high-energy telescopes are one step away from helping with clarifying this puzzle.
\item The large-scale distribution of CR electrons throughout the Milky Way disk (Sec.~\ref{sec:diskmw}) is less constrained by existing observations. However, it is now established thanks to \textit{INTEGRAL}~SPI that there is a diffuse gamma-ray emission above 60~keV consistent with a dominant IC origin, and which suggests abundances of secondary CR electrons and positrons in the inner Galaxy larger than what is observed locally. Improved measurements in the MeV and TeV domains, along with multiwavelength data, are key to improve our understanding of CR electrons.
\item Spatially-resolved gamma-ray observations of the closest external galaxies (Sec.~\ref{sec:localgroup}), namely the LMC, SMC, and M31, reveal a varied picture at GeV energies. Large-scale CR density variations of a factor 2-3 similar to those found in the Milky Way appear in the LMC, but with 50\% of the emission at 10~GeV contributed by extended components of a nature yet to be understood. Extended emission from the SMC lacks remarkable correlation with gas densities and star-formation sites, while in M31 gamma-ray emission is concentrated in innermost 5 kpc, possibly due to the larger concentration of an old stellar population in the bulge. These studies have so far been limited to the GeV range, and the topic has hardly been explored observationally in the TeV range.
\item The \F~LAT has enabled for the first time to image the CR diffusion halo in the Milky Way, and perhaps in M31 (Sec.~\ref{sec:halo}). For the Milky Way observations of gas clouds at large distances from the disk show a marked decline of CR densities at heights above 2 kpc, while it is unclear if large fluctuations found between different clouds at lower altitudes are related to the large-scale vertical gradient of CRs or peculiar transport conditions. These observations complement and inform ongoing theoretical efforts to establish the physical origin of the CR diffusive halo in terms of plasma waves self-excited by the CR streaming instability possibly combined with turbulence advected from the Galactic disk.
\item Standard implementations of CR propagation models for the Milky Way show an overall agreement with observations, but extended residual emission as large as 30\% appears all over the sky in \F-LAT observations (Sec.~\ref{sec:residuals}). Possible explanations include contributions from populations of sub-threshold sources and an incomplete/imperfect census of target gas and radiation fields for gamma-ray production, but part of these residuals probably calls for refinements of the models, such as including a more realistic description of CR sources (e.g., time-dependent particle injection, detailed spatial distribution, including more classes of sources in addition to SNRs) and accounting for a variety of different aspects of transport (e.g., localized confinement in the vicinity of sources, role of galactic winds and outflows).
\item GeV and TeV gamma-ray observations have been unveiling CR propagation in the vicinity of sources (Sec.~\ref{sec:youngcr}) as a field rich in phenomenology, connecting acceleration at the sources and interstellar transport in the non-linear regime. This has the potential to give rise to emission structures and specific populations of gamma-ray sources on various scales. Gamma-ray emission around SNRs can be interpreted in a variety of scenarios: particles just detaching from the shock, escaping particles illuminating nearby clouds, and the reacceleration of old CRs trapped in shocked clouds. Gamma-ray halos surrounding old pulsars are a brand new class of emitters, governed by physics still largely to be understood, and whose long lifetimes may translate into a numerous population yet to be uncovered. Long-suspected gamma-ray emission in the direction of SFRs was finally detected, but without consensus yet on the processes that shape particle acceleration and transport in these environments, and whether they play a distinctive role if the CR life cycle. For most of these cases, theory and observations currently seem to point towards a suppression of the CR diffusion coefficient by one/two order of magnitudes around sources with respect to typical interstellar values, the implications of which deserve further investigation. A robust characterization of CRs in the vicinity of sources needs to go hand-in-hand with the characterization of large-scale interstellar emission and requires an extensive multiwavelength approach.
\item A dozen star-forming galaxies are detected at GeV energies, and two at TeV energies (Sec.~\ref{sec:galaxies}). The data reveal a mildly non-linear correlation with a significance close to 5$\sigma$ between GeV and infrared luminosity. The correlation is driven by massive-star formation, which is at the origin of strong UV/optical light and CRs that, upon interaction with the ISM, power infrared and gamma-ray emission, respectively. The non-linear relation between the two is interpreted as resulting from an increasing CR calorimetric efficiency for galaxies with larger gas and radiation content. As a population, star-forming galaxies are estimated to contribute $\sim$10\% of the extragalactic gamma-ray background, which extrapolates to only $\sim$1\% of the TeV-PeV astrophysical neutrino flux. Several avenues are explored to model CR transport and interactions in a variety of galactic environments, but it is fair to say that our understanding of the CR life cycle still lacks extensive testing for other galaxies. 
\end{itemize}


In the GeV range, where the entire sky was surveyed by the \F~LAT, interstellar emission has been explored in depth.
In the TeV range, however, we have just started to scratch the surface. In the next decade the deep and extensive surveys of the Galactic plane and LMC with CTA, combined with continued operation of HAWC and starting exploitation of LHAASO, and, possibly, the advent fo SWGO (Sec.~\ref{sec:obsgamma}) will unveil what interstellar emission looks like in the TeV range and above. This will shed light on crucial questions such as the origin of the inner-Galaxy hardening. It also has the potential to reveal how interstellar emission from $>$TeV CRs connects to ISM structures and the source regions on a variety of different scales.  Very-high-energy observations are of particular interest to study the early phases of the CR life cycle. In particular, for the study of gamma-ray emission in the source vicinity developments in the very-high-energy range may alleviate the problems encountered when trying to disentangle these objects from large-scale diffuse emission, since we expect that at energies above 100\,GeV large-scale interstellar emission should be comparatively less intense. The good angular resolution of CTA is essential for proper identification of the emission components in complex regions or resolving of fine structures close to the shock in SNRs. At the same time TeV instruments are expected to increase the number of external galaxies detected in gamma rays and provide a broader view of different realizations of the CR phenomenon at very high energies. 

Toward the lower-end of the gamma-ray spectrum, a new (sub-)MeV mission such as ASTROGAM, AMEGO or GECCO (see again Sec.~\ref{sec:obsgamma}) has a huge potential to better reveal spectral and spatial properties of IC emission from the Galaxy and, therefore, to infer the distribution of CR electrons which better sample CR inhomogeneities since they are affected by energy losses more strongly than nuclei and remain much closer to their sources. The improved performance in the MeV to GeV energy range of ASTROGAM or AMEGO would also allow to probe CR nuclei in nearby clouds at sub-pc scale to test how low-energy CRs get depleted at their interior, to follow the release and diffusion of particles around SNRs and pulsars, and to infer the spectral energy distribution of the bulk of CR nuclei in SFRs in order to estimate the CR pressure in these environments. 

Besides developments in gamma-ray instruments, major progress in the years to come is expected to stem from a more extensive integration of gamma-ray observations in a broad multiwavelength/multimessenger context. This is necessary to unambiguously interpret gamma-ray observations, as illustrated along the review, and it is also needed to explore the cross effects with related fields (e.g., star formation, galaxy formation and evolution, astrobiology). 
At the same time, major theoretical and numerical developments are already warranted and under way (Sec.~\ref{sec:modda}) and a challenge to the community is to ensure their open dissemination and use, in order to fully exploit the existing and forthcoming experimental data, and better connect macroscopic observables to the microphysics of non-thermal particle transport.

\vspace{6pt} 



\authorcontributions{The three authors contributed equally to conceptualization, literature review, visualization, and writing.}

\funding{L.T. and P.M. acknowledge financial support by CNES for the exploitation of \F-LAT observation and from CNRS-INSU for their work on CTA, as well as from the French Agence Nationale de la Recherche under reference ANR-19-CE31-0014 (GAMALO project). D.G.\ has received financial support through the Postdoctoral Junior Leader Fellowship Programme from la Caixa Banking Foundation (grant n.~LCF/BQ/LI18/11630014).
D.G. was also supported by the Spanish Agencia Estatal de Investigaci\'{o}n through the grants PGC2018-095161-B-I00, IFT Centro de Excelencia Severo Ochoa SEV-2016-0597, and Red Consolider MultiDark FPA2017-90566-REDC.}

\acknowledgments{The authors acknowledge discussions during the preparation of the review with Peter von Ballmoos, Henrike Fleishack, Michael Kachelrie\ss, Elena Orlando, and Andy Strong. Special thanks to Sarah Recchia for providing the model curve in Figure~\ref{fig:crgrad}, to Carmelo Evoli for providing the model curve in Figure~\ref{fig:vertical_gradient}, to Marco Ajello and Mattia Di Mauro for providing the data in Figure~\ref{fig:gev-correlation}, and to Reshmi Mukherjee for the comments on the manuscript. This work has made use of the SIMBAD database, operated at CDS, Strasbourg, France, and of NASA's Astrophysics Data System Bibliographic Services. The preparation of the figures has made use of the following open-access software tools: APLpy \cite{aplpy}, Astropy \cite{astropy2013}, Matplotlib \cite{matplotlib}, NumPy \cite{numpy}, SciPy \cite{scipy}.}

\conflictsofinterest{The authors declare no conflict of interest. The funders had no role in the design of the review, in the literature selection, in the writing of the manuscript, or in the decision to publish.} 



\abbreviations{The following abbreviations are used in this manuscript:\\

\noindent 
\begin{tabular}{@{}ll}
CDS & Centre de Donn\'ees astronomiques de Strasbourg \\
CR & cosmic ray\\
GC & Galactic center\\
HE & high energy\\
IC & inverse Compton \\
ISM & interstellar medium\\
LAT & Large Area Telescope \\
LIS & local interstellar spectrum\\
LMC & Large Magellanic Cloud\\
PSF & point spread function\\
PWN & pulsar wind nebula\\
SBG & starbust galaxy \\
SFG & star-forming galaxy \\
SFR & star-forming region\\
SMC & Small Magellanic Cloud\\
SNR & supernova remnant\\
VHE & very-high energy
\end{tabular}}

%
%
%
\end{paracol}
\reftitle{References}


\externalbibliography{yes}
\bibliography{references.bib,CosmicRayEscape.bib,StarFormingRegions.bib,Halo.bib,LMC.bib,Starbursts.bib}

\end{document}